\documentclass[useAMS,usenatbib]{mn2e}
\usepackage{graphicx}
\usepackage[figuresright]{rotating}

\def\ic{IC\,2944}

\def\cpd{CPD\,$-$62$^{\circ}$}

\def\l{~$\lambda$}
\def\ll{~$\lambda\lambda$}

\def\hb{H\,{\sc ii}}
\def\hea{He\,{\sc i}}
\def\heb{He\,{\sc ii}}
\def\nc{N\,{\sc iii}}
\def\nd{N\,{\sc iv}}
\def\ne{N\,{\sc v}}
\def\mgb{Mg\,{\sc ii}}

\def\sid{Si\,{\sc iv}}

\def\kms{km\,s$^{-1}$}

\def\msol{M$_{\odot}$}
\def\rsol{R$_{\odot}$}

\def\Rsun{R$_{\odot}$}
\def\s{$\sigma$}
\def\w{$\omega$}
\def\vsini{$v \sin i$}

\def\feros{{\sc feros}}

\def\uves{{\sc uves}}

\def\na{{\it n/a}}

\def\sss{\subsubsection}

\defcitealias{SBL98}{SBL98}
\defcitealias{GM01}{GM01}
\defcitealias{HCB74}{HCB74}
\defcitealias{LM83}{LM83}
\defcitealias{PHYB90}{PHYB90}
\defcitealias{PHC91}{PHC91}
\defcitealias{BL95}{BL95}
\defcitealias{RCB97}{RCB97}
\defcitealias{BVF99}{BVF99}
\defcitealias{SGN08}{Paper~I}
\defcitealias{SGE09}{Paper~II}
\defcitealias{ESL05mnras}{ESL05}
\defcitealias{GPM07}{GPM07}

\title[The O star binary fraction in IC~2944 and Cen~OB2]{The massive star binary fraction in young open clusters - III. IC 2944 and the Cen OB2 association}
\author[H. Sana et al.]{H. Sana$^{1,2}$\thanks{E-mail: H.Sana@uva.nl}, 
   G. James$^{3}$ and
   E. Gosset$^{4}$\thanks{F.R.S. - FNRS, Belgium}\\
$^{1}$Sterrenkundig Instituut Anton Pannekoek, Universiteit van Amsterdam, Science Park 904, 1098 XH Amsterdam, The Netherlands\\
$^{2}$European Southern Observatory, Alonso de Cordova 1307, Casilla 19001, Santiago 19, Chile\\
$^{3}$European Southern Observatory, Karl-Schwarzschild-Strasse 2, D-85748 Garching bei M\"unchen, Germany\\
$^{4}$Astrophysical Institute, Li\`ege University, B\^at. B5c, All\'ee du 6 Ao\^ut 17, B-4000 Li\`ege, Belgium
}

\begin{document}

\date{Accepted 1988 December 15. Received 1988 December 14; in original form 1988 October 11}

\pagerange{\pageref{firstpage}--\pageref{lastpage}} \pubyear{2002}

\maketitle

\label{firstpage}

\begin{abstract}
Using an extended set of multi-epoch high resolution high signal-to-noise ratio optical spectra, we readdress the multiplicity properties of the O-type stars in IC~2944 and in the Cen OB2 association.  We present new evidence of binarity for five objects and we confirm the multiple nature of another two. We derive the first orbital solutions for HD~100099, HD~101436 and HD~101190 and we provide additional support for HD~101205 being a quadruple system. The minimal spectroscopic binary fraction in our sample is $f_\mathrm{min}=0.57$. Using numerical simulations, we show that the detection rate of our observational campaign is close to 90\%, leaving thus little room for undetected spectroscopic binary systems.  The statistical properties of the O-star population in IC~2944 are similar, within the uncertainties, to the results obtained in the earlier papers in this series despite the fact that sample size effects limit the significance of the comparison. Using newly derived spectroscopic parallaxes, we reassess the distance to IC~2944 and obtained $2.3~\pm0.3$~kpc, in agreement with previous studies. We also confirm that, as far as the O stars are concerned, the IC~2944 cluster is most likely a single entity.
\end{abstract}


%***************************************************************
%***********  Measured spectral lines **************************
%*************************************************************** 
\begin{table*}
 \begin{minipage}{170mm}
  \caption{Number of radial velocity (RV) measurements obtained for each object with respect to the considered spectral lines. A `--' means that the corresponding line has not been measured. } \label{tab: sp_lines}
 \centering
  \begin{tabular}{@{}rrrrrrrrrrrrr@{}}
  \hline
         & \hea       & \sid       & \hea       & \hea       & \hea       & \mgb       & \heb       & \heb       & \hea     & \heb       & \hea       & \hea     \\
Object   & \l4026     & \l4089     & \l4144     & \l4388     & \l4471     & \l4481     & \l4542     & \l4686     & \l4922   & \l5412     & \l5876     & \l7065   \\
\hline     
HD~100099 &  24       & 24         & 24         & 24         & 24         & 24         & 24         & 24         & 24       & 14         & 24         & 24       \\ %done
%HD~101131 &   1       &  1         & 1          & 1          & 1          & 1          & 1          & 1          & 1        & 1          & 1          &  1       \\
HD~101190 &  20       & 20         & 20         & 20         & 20         &  --        & 20         & 20         & 20       & 10         & 20         &  20      \\ %done
HD~101191 &  12       & 12         & 12         & 12         & 12         & 12         & 12         & 12         & 12       &  7         & 12         &  12      \\ %done    
%HD~101205 &  11       & 11         & 10         & 12         & 12         & 10         & 12         & 13         & 9        & 10         & 10         &  9       \\
HD~101223 &   6       &  6         &  6         &  6         &  6         &  6         &  6         &  6         &  6       &  4         &  6         &  6       \\%done
HD~101298 &   8       &  8         &  4         &  8         &  8         &  8         &  8         &  8         &  8       &  6         &  8         &  8       \\%done

HD~101333 &   5       &  5         &  5         &  5         &  5         &  5         &  5         &  5         &  5       &  3         &  5         &   5      \\%done
HD~101413 &  10       & 10         & 10         & 10         & 10         & 10         & 10         & 10         & 10       &  5         & 10         &  10      \\%done
HD~101436 &  13       & 13         & 13         & 13         & 13         &  --        & 13         & 13         & 13       &  8         & 13         &  13      \\%done
HD~101545A&   7       &  7         &  7         &  7         &  7         &  7         &  7         &  7         &  7       &  5         &  7         &  7       \\%done
HD~101545B&   7       &  7         &  7         &  7         &  7         &  7         &  7         &  7         &  7       &  5         &  7         &  7       \\%done
HD~308804 &   2       &  --        &  2         &  2         &  2         &  2         &  --        &  --        &  2       &  --        &  2         &  --      \\%done
HD~308813 &  11       & 11         & 11         & 11         & 11         & 11         & 11         & 11         & 11       &  6         & 11         &  11      \\%done
\cpd2198  &   6       &  6         &  6         &  6         &  6         &  5         &  6         &  6         &  6       &  4         &  6         &  6       \\ %done
\hline
\end{tabular}
\end{minipage}
\end{table*}
%***************************************************************

\section{Introduction}
Best known for its spectacular group of globules \citep{Tha50, RCO97, RRH03}, \ic\ is both an \hb\ region and a rich stellar complex at the inner edge of the Carina spiral arm.  Although the stellar aggregate formed by the bright OB stars in the vicinity of HD~101205  \citep[O7~IIIn((f)), ][]{Wal73} is located closer to the IC~2948 \hb\ region, \citet{Col31} catalogued it as \ic. The same global group of stars was further listed as the Cen OB2 association by \citet{ABR70}. The whole complex is embedded in the large ionization region  RCW~62, that includes  the smaller \ic\ and IC~2948 \hb\ regions and extends over about one square-degree.  Some authors \citep[e.g.,][]{KPR05} also distinguished the separate stellar cluster IC~2948, which is formed by the few bright stars around HD~101413, at $\sim12$\arcmin\ S-E of HD~101205, and placed both entities at different distances. As a result, these stars have double entries in the stellar cluster data base WEBDA \citep{Me92}. In this paper, we follow the earlier approach of \citet{Col31} and \citet{ArM77} and we use the IC~2944 denomination for the complete stellar aggregate at the core of RCW~62. The homogeneity of the region and its distance will be discussed later in this paper.

\citet{ThW65} performed the first extensive photometric and spectroscopic study of the cluster. They identified 14 early B- and nine O-stars, and deduced  a distance $d\approx2.0\pm0.2$~kpc. Several authors subsequently questioned the physical  reality of the cluster. \citet{ArM80} suggested that the stellar aggregate was rather formed by several distinct stellar groups, with distances ranging from 0.7 to 4~kpc. \citet{PeL86} concluded that the apparent stellar concentration of \ic\ resulted from the superposition of isolated early-type stars along the line of sight. \citet{Wal87} reaffirmed, as far as the O-type stars are concerned, the physical reality of the cluster and showed that the spectroscopic parallax of the O-type stars was in excellent agreement with the earlier distance proposed by \citet{ThW65}. Polarimetric observations by \citet{VOM94} confirmed that some of the groups identified by \citet{ArM80} showed clear differences while others could not be separated from each other. The more recent study of \citet{TEH98}, based on the UV space telescope {\sc glazar}  on board the Mir space station \citep{TKK88}, reassessed the situation. Studying 185 OBA stars in an area of 8 deg$^2$ towards Cen OB2, they confirmed that most of the O stars belong to \ic, at a distance of $2.2\pm0.3$~kpc. They further identified five other OB associations with distances ranging from 0.85 to 6.7~kpc (see their Table 5).

As the third paper in our series, the present work re-address the spectral and multiplicity properties of the  O-type star population in Cen~OB2 and IC~2944. It is organized as follows. Sect.~\ref{sect: obs} summarizes the observing campaign and the data reduction. Sect.~\ref{sect: ostar} discusses the individual objects. Sect.~\ref{sect: MC} evaluates the observational biases  while Sect.~\ref{sect: discuss} describes our results and Sect.~\ref{sect: ccl} summarizes our work.

%***************************************************************
%************ Journal of the observations   ********************
%***************************************************************
\begin{sidewaystable*}
\begin{minipage}[t][180mm]{\textwidth}\begin{flushleft}
{\bf Table 2.} ~Journal of the spectroscopic observations of the O-type stars in IC\,2944. First and second lines indicate the spectral line and the adopted rest wavelength (in \AA). The first column gives the heliocentric Julian date at mid-exposure. The following columns provide, for each spectral line, the measured RVs (in \kms).  References for the instrumental setup can be found at the bottom of the table. The full table is available in the electronic edition of the journal.\end{flushleft}
 \centering
% \label{tab: diary}
\begin{tabular}{@{}rrrrrrrrrrrr@{}}
\hline												
HJD           & \hea\l4026 & \sid\l4089 & \hea\l4144 &\hea\l4388 & \hea\l4471 & \mgb\l4481 & \heb\l4542 & \heb\l4686 & \hea\l4922 &\heb\l5412 &\hea\l5876\\
$-2\,400\,000$& 4026.072   &  4088.863  &  4143.759  & 4387.928  &  4471.512  &  4481.228  &  4541.590  &  4685.682  &  4921.929  & 5411.520  & 5875.620 \\
\hline 
\multicolumn{12}{c}{HD~100099 prim}\\
\hline
  53135.461$^{a}$ &   $-$8.89 & $-$16.29 & $-$13.65 & $-$12.42 & $-$14.85 & $-$14.58 & $-$14.50 & $-$10.13 & $-$15.16 & $-$14.97 & $-$13.78 \\
  53509.559$^{a}$ &  $-$74.59 & $-$75.76 & $-$74.01 & $-$73.28 & $-$75.24 & $-$74.77 & $-$73.25 & $-$75.65 & $-$75.47 & $-$74.81 & $-$75.03 \\
  53511.507$^{a}$ &  $-$14.22 & $-$20.77 & $-$15.12 & $-$14.20 & $-$17.73 & $-$20.62 & $-$22.40 & $-$15.43 & $-$20.55 & $-$21.60 & $-$21.42 \\
  53512.476$^{a}$ &     26.71 &    30.95 &    25.39 &    24.05 &    25.36 &    25.56 &    31.70 &    30.56 &    26.32 &    30.49 &    28.49 \\
  53860.551$^{a}$ &    145.89 &   149.61 &   144.70 &   143.07 &   143.97 &   143.91 &   146.30 &   149.39 &   143.75 &   148.34 &   147.65 \\
  53860.704$^{a}$ &    142.04 &   147.69 &   140.38 &   140.81 &   141.98 &   141.91 &   142.84 &   147.62 &   141.93 &   143.66 &   144.56 \\
  53861.510$^{a}$ &    115.26 &   119.97 &   114.35 &   116.25 &   115.79 &   115.79 &   118.31 &   122.67 &   116.40 &   119.63 &   119.18 \\
  53861.688$^{a}$ &    110.78 &   116.61 &   112.14 &   112.17 &   111.84 &   111.84 &   113.11 &   116.74 &   110.88 &   114.14 &   112.49 \\
  53862.485$^{a}$ &     91.09 &    89.30 &    88.28 &    87.57 &    87.73 &    87.73 &    89.40 &    94.24 &    87.17 &    90.65 &    93.88 \\
  53862.681$^{a}$ &     79.54 &    83.59 &    79.65 &    79.36 &    79.66 &    79.72 &    83.57 &    86.23 &    79.87 &    83.49 &    84.69 \\
  53863.489$^{a}$ &     52.75 &    56.45 &    53.61 &    55.87 &    53.50 &    53.64 &    61.38 &    61.31 &    54.24 &    60.40 &    58.69 \\
  53863.730$^{a}$ &     53.50 &    48.27 &    47.10 &    46.63 &    45.43 &    45.64 &    53.51 &    52.36 &    50.40 &    52.10 &    51.00 \\
  53864.501$^{a}$ &     32.62 &    32.04 &    31.92 &    32.27 &    31.35 &    31.62 &    39.63 &    38.28 &    32.56 &    38.77 &    32.93 \\
  53864.727$^{a}$ &     28.18 &    29.07 &    27.61 &    28.15 &    27.35 &    27.56 &    33.71 &    34.45 &    26.92 &    33.39 &    27.91 \\
  54493.820$^{b}$ &   $-$7.53 & $-$10.13 &  $-$2.45 &  $-$6.61 & $-$10.22 &  $-$9.88 &  $-$9.94 &  $-$4.43 &  $-$9.98 &       -- & $-$10.07 \\
  54513.845$^{b}$ &      6.62 &     6.92 &     8.44 &     6.43 &     5.90 &     6.17 &     5.91 &    10.04 &     6.41 &       -- &     6.62 \\
  54514.883$^{b}$ &   $-$1.89 &  $-$3.10 &     2.23 &  $-$1.64 & $-$11.09 &  $-$5.39 &  $-$5.64 &     0.85 &  $-$1.67 &       -- &  $-$3.03 \\
  54515.814$^{b}$ &   $-$8.62 & $-$12.93 & $-$12.48 & $-$11.01 & $-$12.50 &  $-$9.43 &  $-$8.93 &  $-$5.57 & $-$12.33 &       -- & $-$12.00 \\
  54521.779$^{b}$ &  $-$76.11 & $-$76.10 & $-$75.95 & $-$75.72 & $-$75.58 & $-$76.18 & $-$75.98 & $-$73.92 & $-$75.82 &       -- & $-$76.08 \\
  54532.799$^{b}$ &     32.60 &    37.83 &    33.00 &    33.24 &    34.59 &    33.58 &    33.19 &    39.21 &    33.28 &       -- &    38.04 \\
  54537.857$^{b}$ &  $-$10.93 & $-$15.73 & $-$11.57 & $-$11.88 & $-$18.57 & $-$14.02 & $-$19.16 & $-$12.28 & $-$16.41 &       -- & $-$18.91 \\
  54539.811$^{b}$ &  $-$39.02 & $-$39.85 & $-$38.62 & $-$39.48 & $-$45.75 & $-$45.35 & $-$45.88 & $-$40.94 & $-$44.74 &       -- & $-$45.66 \\
  54540.817$^{b}$ &  $-$50.82 & $-$51.23 & $-$51.31 & $-$51.47 & $-$55.34 & $-$58.09 & $-$55.58 & $-$52.26 & $-$52.34 &       -- & $-$57.31 \\
  54544.539$^{b}$ &  $-$76.68 & $-$77.59 & $-$76.09 & $-$76.85 & $-$78.97 & $-$79.50 & $-$75.40 & $-$72.94 & $-$78.83 &       -- & $-$79.38 \\
\hline
\end{tabular}\\
\begin{flushleft}$a.$ ESO2.2m + FEROS ; $b.$ VLT + UVES
\end{flushleft}
\end{minipage}
\end{sidewaystable*}
%***************************************************************
%***************************************************************
%***************************************************************

% =================== EW for each object ===================================

\setcounter{table}{+2}
\begin{table*}
 \centering
 \begin{minipage}{150mm}
  \caption{Diagnostic line ratios and corresponding spectral types for the studied O-type objects. Inapplicable criteria for specific cases are marked as \na. We refer to text for discussion of the individual cases and to Table~\ref{tab: bin} for the finally adopted spectral types. Quoted error-bars give the 1-\s\ dispersions on the measurements.} 
\label{tab: EW}
  \begin{tabular}{@{}llrrrrl@{}}
  \hline
Object     & Component     & $\log W'$        & $\log W''$      &$\log W(\lambda4686)$& $\log W'''$     &  Sp. Type\\
  \hline     
HD~100099 & prim           &$0.305\pm0.066$   & $0.292\pm0.021$ & \na                & $>4.944\pm0.016$ & O9 III/I  \\
HD~100099 & sec            &$0.660\pm0.067$   & $-0.117\pm0.054$& \na                & $>4.361\pm 0.062$& O9.7 V    \\
%HD~101131 & 		   & 		      & 		& 		     & 	               & 	\\
HD~101190 & composite	   & $-0.369\pm0.023$ & \na		& $2.737\pm0.028$    & \na	       & O5.5~V	   \\
HD~101191 & composite	   & $0.131\pm0.006$  & $0.102\pm0.032$	& $2.864\pm0.005$    & \na             & O8 V/III  \\
%HD~101205 & prim	   & 		      & 		& 		     & 	               & 	   \\
%HD~101205 & sec 	   & 		      & 		& 		     & 	               & 	   \\
HD~101223 & \na            & $ 0.148\pm0.022$ & $0.192\pm0.054$ &    $2.777\pm0.026$ & \na             & O8 V      \\
HD~101298 & \na            & $-0.200\pm0.035$ & $0.423\pm0.143$ &    $2.705\pm0.037$ & \na             & O6-O6.5 V/III \\
HD~101333$^a$ & \na	   & $0.315-0.666$    & $0.118-0.144$	& $5.25-5.69$	     & \na             & O9-O9.7 V/III \\
HD~101413 & prim	   & $0.112\pm0.017$  & $-0.038\pm0.021$& $>2.793\pm0.013$   & \na             & O8 V	   \\
HD~101436 & prim	   & $-0.090\pm0.050$  & $0.302\pm0.230$	& $>2.061\pm0.147$   & \na             & O6.5-O7 V\\
HD~101436 & sec		   & $-0.170\pm0.107$  & \na		& $>2.761\pm0.062$   & \na             & O5-O6 V  \\
HD~101545 & A              & $ 0.461\pm0.041$ & $0.282\pm0.048$ & \na                & $5.082\pm0.067$ & O9.5 III \\
HD~101545 & B              & $ 0.890\pm0.179$ & $0.234\pm0.094$ & \na                & $4.950\pm0.060$ & O9.7 III/I\\
HD~308813 & prim	   & $ 0.599\pm0.025$ & $0.015\pm0.104$	& \na		     &$>5.378\pm0.038$ & O9.5 V       \\
\cpd2198  & \na            & $0.663\pm0.053$  & $0.205\pm0.083$ & \na                & $5.110\pm0.053$ & O9.7 III/I\\
\hline
\end{tabular}\\
$a.$ The nature of the adopted line profile is the dominant source of uncertainties, so that we rather quote here the range of values obtained by fitting either Lorentz or Gaussian profiles. \\
\end{minipage}

\end{table*}
% ========================================================================

%%%%%%%%%%%%%%%%%%%%%%%%%%%%%%%%%%%%%%%%%%%%%%%%%%%%%%%%%%%%%%%%%%%%%
%%%%%%%%%%%%%%%%%%%%%%%%%%%%%%%%%%%%%%%%%%%%%%%%%%%%%%%%%%%%%%%%%%%%%

\section{Observations and data handling} \label{sect: obs}
This paper is based on 166 high-resolution echelle spectra covering most of the optical spectrum of the targets. 97 of those spectra were obtained from 2004 to 2007  with the \feros\ spectrograph at the ESO~2.2m telescope (La Silla, Chile) with a resolving power of 48,000. The other 69 spectra were taken from Jan to Mar 2008 using the \uves\ spectrograph at the ESO VLT/UT2 telescope (Paranal, Chile) operated in dichroic mode with the DIC2 437+760 setup and an 0.8\arcsec\ entrance slit for both arms. The resulting resolving power is 50,000, thus almost identical to the \feros\ one.  The data reduction is similar to the one presented in \citet[][\citetalias{SGE09}]{SGE09} and the description will not be repeated here. The final, one-dimensional spectra were normalized to unity before merging the individual orders. The average signal-to-noise ratio (SNR) of the data set is well above 150. 

Line profile fitting was performed with a semi-interactive code ({\tt fitline}) developed by P. Fran{\c c}ois at the Paris-Meudon Observatory and originally used in the context of chemical abundance determinations in globular cluster stars \citep{JFB04} or in extremely metal-poor halo stars \citep{FDH07}. This code is based on a genetic algorithm approach \citep{Cha95} that mimics the way genetic mutations affect DNA, driving the evolution of species under given environmental constraints. In our case, lines are fitted by  Gaussian profiles, each of them defined by four parameters: the central wavelength, the width and depth of the line, and the continuum value. Briefly, the algorithm proceeds as follows: 

\begin{enumerate}
 \item An initial set of Gaussians is computed, giving random values to the four parameters. The quality of each fit is estimated by a $\chi^2$ calculation.

\item A new ``generation'' of Gaussians is then calculated from the 20 best fits after adding random modifications to the initial set of parameters (``mutations''). The new set replaces the old one, new fits are computed and the quality of these fits is estimated again by using a $\chi^2$ evaluation.

\item The process is iterated typically several hundred times until convergence to the best Gaussian fits is attained.

\item A line-by-line interactive inspection of the fits is performed to remove bad fits, to correct the position of the continuum, or to handle line blends in SB2 systems.

\item The process is iterated again, and the profile parameters corresponding to the best fit are adopted.
\end{enumerate}

The spectral lines considered in this process are listed in Table \ref{tab: sp_lines} together with the number of measurements available for each star. The equivalent width (EW) and the radial velocity (RV) of each line are computed based on the final fit, using the same rest wavelengths as adopted in other papers of this series. The rest wavelengths are further indicated in Table~2 that provides the journal of the observations and that lists,  line-by-line, the RV measurements.

As in other papers in the series, we adopted the spectral classification criteria of \citet{Ca71}, \citet{Con73_teff}, \citet{Mat88} and \citet{Mat89}, that  rely on the EWs of given diagnostic lines. We used the usual notations: $\log W'= \log W(\lambda4471) - \log W(\lambda4542)$,  $\log W''= \log W(\lambda4089) - \log W(\lambda4144)$ and  $\log W'''= \log W(\lambda4388) + \log W(\lambda4686)$, where the equivalent widths are expressed in m\AA. The measurements and the corresponding spectral types are given in Table \ref{tab: EW} and discussed in Sect.~\ref{sect: ostar}.  The EWs of SB2 lines were measured at epochs of large separation only. Table~\ref{tab: EW} provides the actual observed values. For multiple systems, because of dilution by the companion continuum, the tabulated values of $\log W(\lambda4686)$ and of $\log W'''$ should be regarded as lower limits.

In Sects.~\ref{sect: SB2} and \ref{sect: SB1}, time series are searched for periodicities using the  Fourier analysis of \citet{HMM85} as amended by \citet{GRR01}.  Quoted uncertainties on the period values are adopted to correspond to one tenth of the width of the associated peak in the periodograms. Whenever relevant, orbital solutions are adjusted using the Li\`ege orbital solution package \citep[e.g.,][]{SGR06_219}.

%%%%%%%%%%%%%%%%%%%%%%%%%%%%%%%%%%%%%%%%%%%%%%%%%%%%%%%%%%%%%%%%%%%%%
%%%%%%%%%%%%%%%%%%%%%%%%%%%%%%%%%%%%%%%%%%%%%%%%%%%%%%%%%%%%%%%%%%%%%
\section{O-type stars in IC~2944} \label{sect: ostar}

%%%%%%%%%%%%%%%%%%%%%%%%%%%%%%%%%%%%%%%%%%%%%%%%%%%%%%%%%%%%%%%%%%%%%

% ========================================================================
\begin{figure}
\centering
\includegraphics[width=\columnwidth]{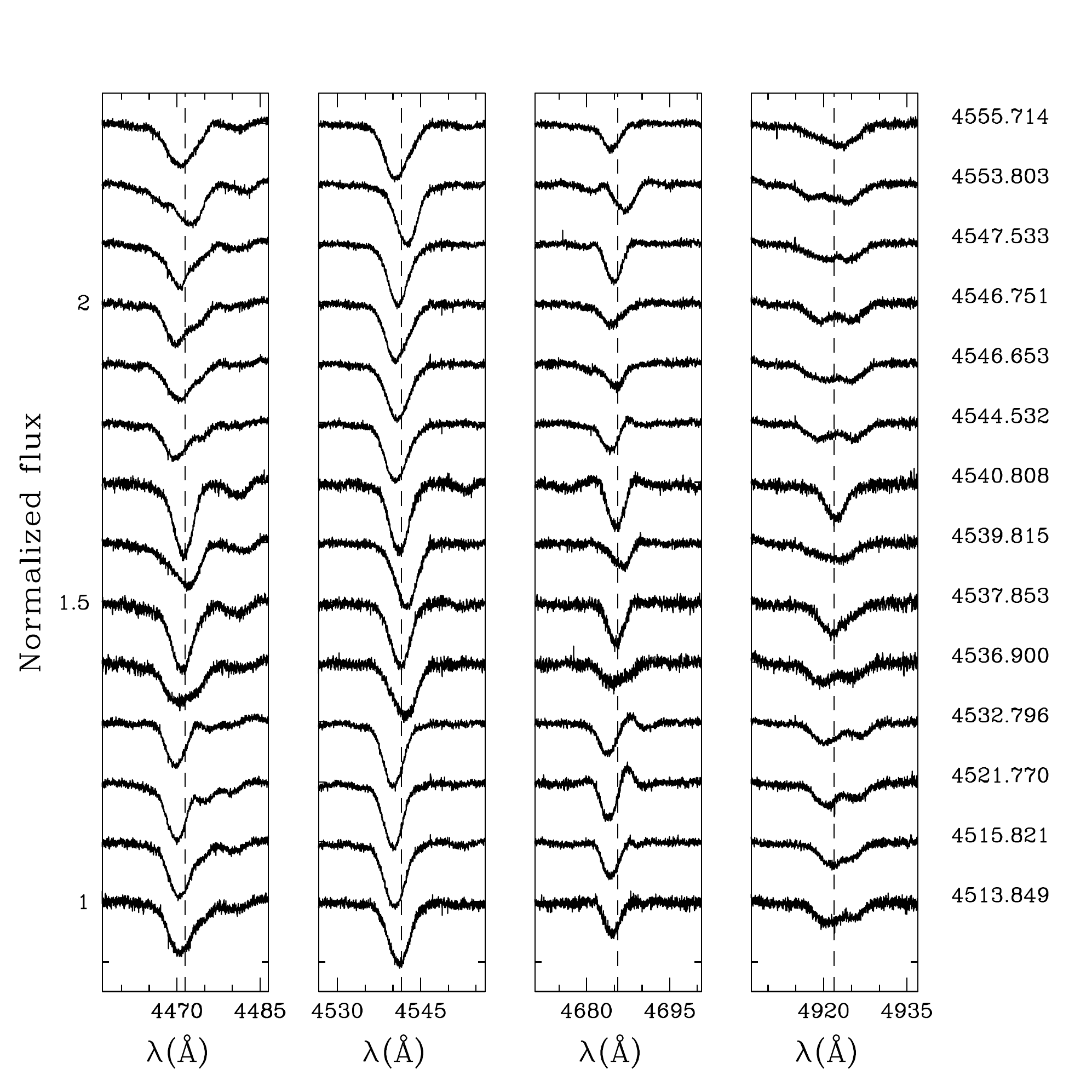}
\caption{{\bf HD~101205:} \hea\l4471, \heb\ll4542, 4686 and \hea\l4922 line profiles at various epochs. The right-hand column gives the heliocentric Julian date (HJD) at mid-exposure in format HJD$-$2\,450\,000. The vertical dashed lines indicate the rest wavelength.}
\label{fig: hd205}
\end{figure}
% ========================================================================

\subsection{SB2 systems}\label{sect: SB2}
\sss{HD~101131}
Close to the cluster core, HD~101131 is a known SB2 binary with O6.5~V((f)) and O8.5~V components. \citet{GPM02} published a detailed analysis of the orbital properties. They derived a period $P$ of 9.6 days and a slight eccentricity $e=0.17\pm0.03$. Because HD~101131 is already well constrained, we did not reobserve it.

\sss{HD~101205}
With a combined magnitude of $V=6.45$, HD~101205 is a visual binary with a separation of 0.36\arcsec\ and $\Delta V=0.3$ \citep{MGH98}. It is also the brightest source in the cluster and a known eclipsing system with a period of 2.1~d \citep{Bal92, MLD92, Ote07}. Additional photometry \citep{MBL10} reveals however a more complicated light curve than expected for an eclipsing binary suggesting that the true nature of the system is not fully understood. We have acquired 20 FEROS spectra from 2004 to 2006 and 15 UVES spectra from Jan to Mar 2008. Our data reveal a clear multiple signature in the \hea\ lines and an SB1 motion and asymmetric profiles of the \heb\ photospheric lines (Fig.~\ref{fig: hd205}). \heb\l4686 presents a mixed absorption and emission profile, with the main absorption component moving in phase with the primary and, possibly, a second fainter absorption component associated to the secondary. In the latter case, the third, overimposed emission component could be produced by wind-wind collision in such a short period system, a situation similar to HD~152248 \citep{SRG01}.

The highest peak in the periodogram corresponds to a period of 2.8~d, a value significantly offset from the photometric period. Adopting this period, we have been able to fit the RVs using a circular orbital solution
(as expected for such a short period system),  yet the residuals are widely spread: 18~\kms. This is significantly larger than the expected RV accuracy ($<$ 5~\kms). This suggests either a systematic error in the measurements (e.g., the presence of a third spectroscopic component), or an additional variability signal.  Naturally, the presence of the speckle companion implies that a third light spectral component must be present, and line blending would tend to reduce the measured velocity excursions. It is unclear whether it would however bias the periodicity of the signal. Finally, a multi-periodic analysis of the time series reveals however no extra-periodic signal on the time scales covered by our campaign, beside the 2.8~d period. 

Our data set is unfortunately insufficient to solve the discrepancy between the photometric and spectroscopic periods. As noted by \citet{MBL10}, the two visual components of HD~101205 might both be multiple systems. A dedicated monitoring is needed to elucidate the nature of this object, which could be a new Trapezium system. The multiple nature of HD~101205 is, however, no longer in doubt.

%Difficult eclipsing system. Clear period of the order of 2.5-3 days.
%\citet{MGH98} quoted a 6.5 magnitude, but noted the presence of closed by companion, thus bringing the magnitude to 7.1 instead. 
%\citet{MGH98} reported three nearby companion at respectively 0.36, 1.7 and 9.6\arcsec\ and $V$ magnitude 7.4, 10.3 and 12.0. 

% ========================================================================
\begin{figure}
\centering
\includegraphics[width=\columnwidth]{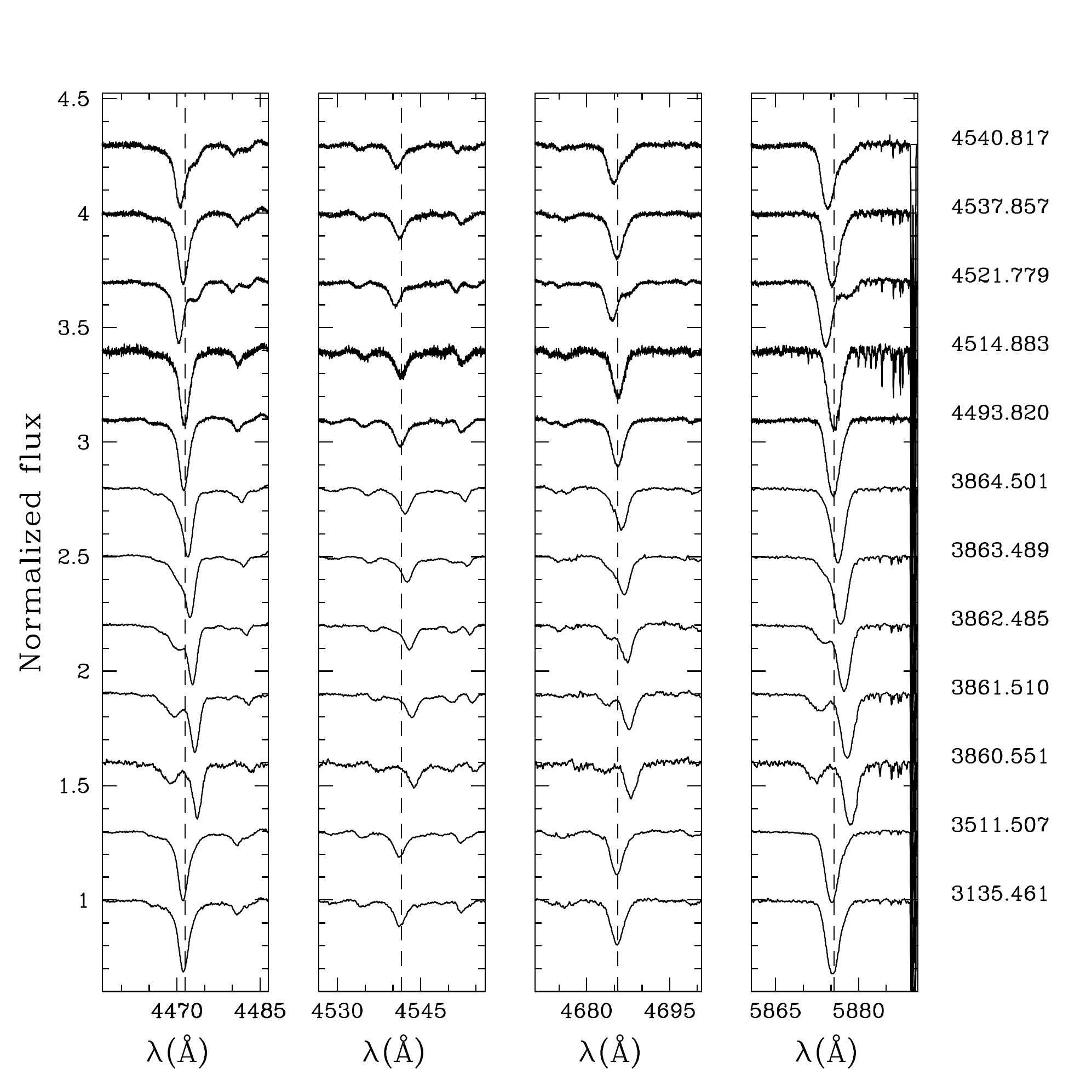}
\caption{{\bf HD~100099:} \hea\l4471, \heb\ll4542, 4686 and \hea\l5876 line profiles at various epochs.}
\label{fig: hd099}
\end{figure}
% ========================================================================

% ========================================================================
\begin{table}
 \begin{minipage}{80mm}
   \centering
   \caption{{\bf HD~100099:} best-fit orbital solution based on the \hea\l5876 RV measurements. $T$  (in HJD$-$2\,450\,000) is the time of periastron passage and is adopted as $\phi=0.0$ in Fig. ~\ref{fig: hd099os}. Quoted uncertainties correspond to 1-\s\ error-bars.}
   \label{tab: hd099os}
   \begin{tabular}{@{}lll@{}}
     \hline
Parameter              & \hea\l5876    \\           
  \hline
$P$ (d)                & $21.5585\pm0.0330$\\ 
$e$                    & $0.517\pm0.019$   \\ 
\w\ (\degr)            & $305.5\pm0.9$     \\ 
$T$                    & $2995.654\pm0.078$\\ 
$\gamma_1$ (\kms)      & $16.5\pm2.1$      \\ 
$\gamma_2$ (\kms)      & $-6.0\pm2.6$      \\ 
$K_1$ (\kms)           & $138.4\pm5.7$     \\ 
$K_2$ (\kms)           & $174.8\pm7.3$     \\ 
 $M_2/M_1$             & $0.792\pm0.013$   \\ 
$a_1 \sin i$ (\rsol)   & $50.46\pm2.20$    \\ 
$a_2 \sin i$ (\rsol)   & $63.72\pm2.78$    \\ 
$M_1\sin^3i$ (\msol)   & $24.0\pm3.1$      \\ 
$M_2\sin^3i$ (\msol)   & $19.0\pm2.4$      \\ 
rms (\kms)             & $2.4$             \\ 
\hline
\end{tabular}
\end{minipage}
\end{table}
% ========================================================================

% ========================================================================
\begin{figure}
\centering
\includegraphics[width=\columnwidth]{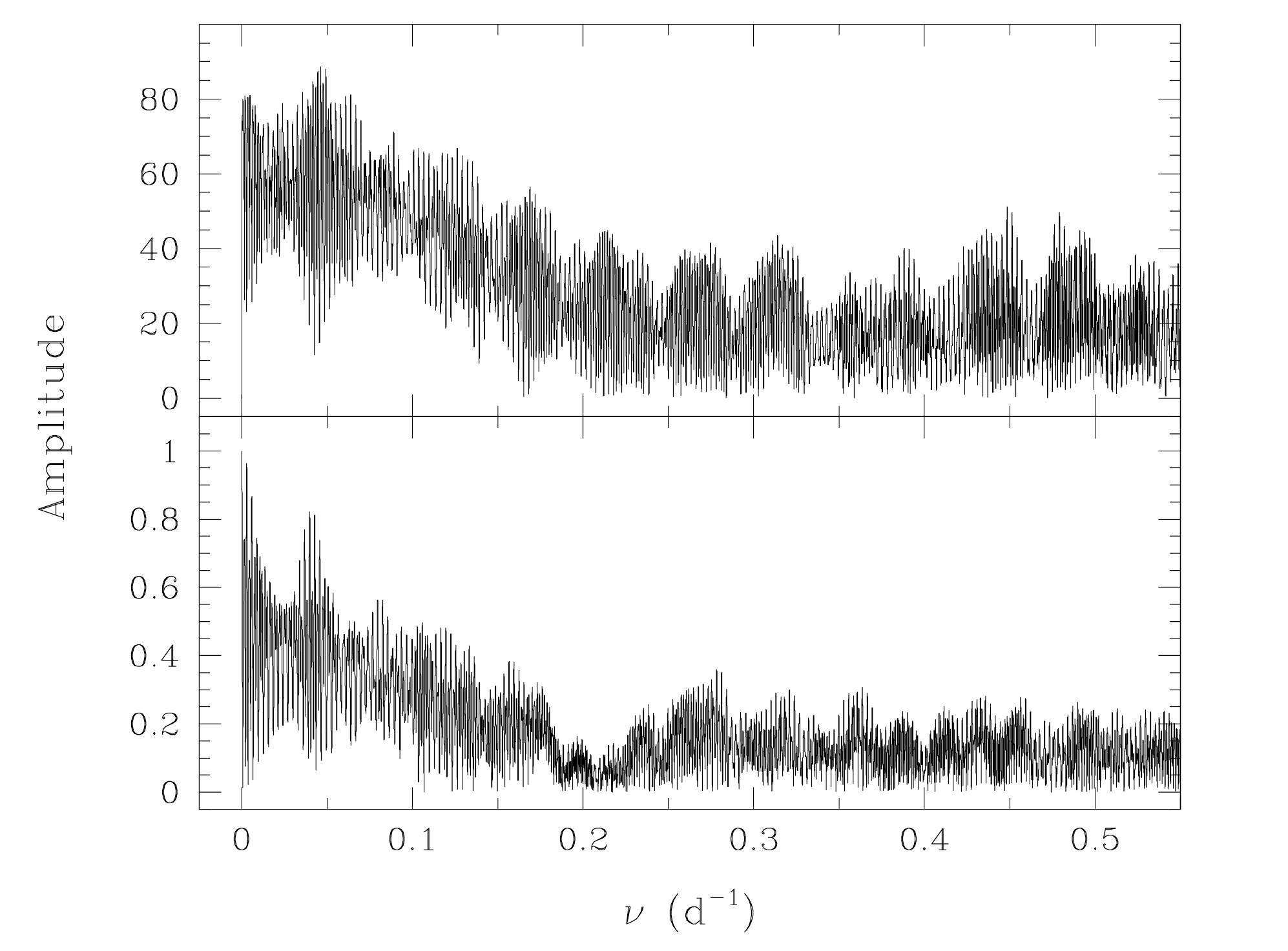}
\caption{{\bf HD~100099:} periodogram (upper panel) and spectral window (lower panel) computed from the \hea\l5876 primary RV measurements.}
\label{fig: hd099period}
\end{figure}
% ========================================================================
% ========================================================================
\begin{figure}
\centering
\includegraphics[width=\columnwidth]{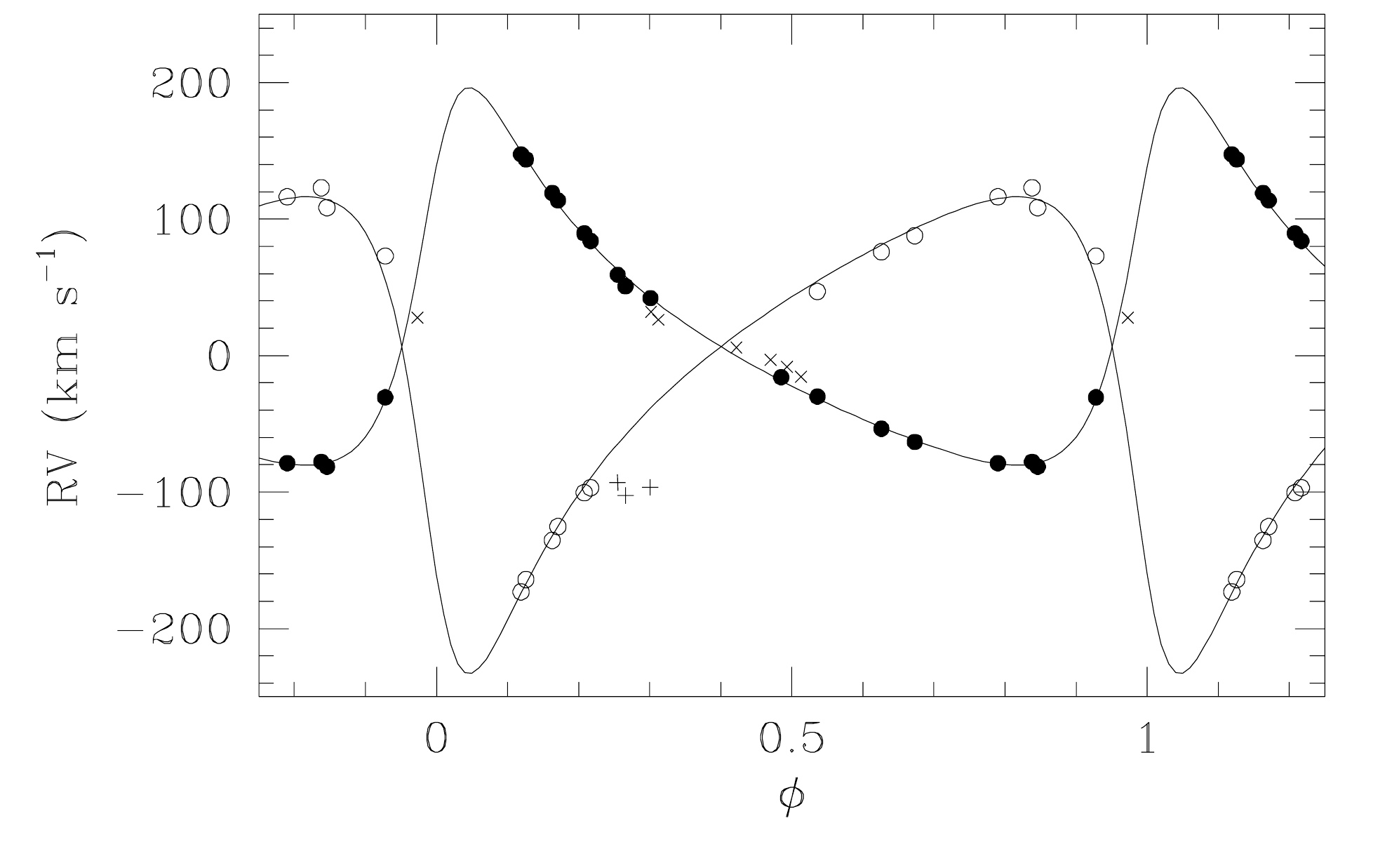}
\caption{{\bf HD~100099:} primary (filled symbols) and secondary (open symbols) RVs together with the best-fit RV curves. The crosses indicate the data not taken into account in the fit because components could not be securely separated. }
\label{fig: hd099os}
\end{figure}
% ========================================================================

\sss{HD~100099}
At about 1\degr\ SW of HD~101205, HD~100099 is associated to IC~2944 in the earlier works \citep[e.g.,][]{ThW65} although \citet{TEH98} preferred to assign it to the Cen-Cru OB2.7 association. Based on the large angular separation of HD~100099  from the cluster center, \citet{BDW00} assigned to it a low membership probability but noted that its proper motion was still compatible with the kinematic properties of IC 2944.

HD~100099 was classified O9~III by \citet{GHS77} and by \citet{Her75}. We collected 24 spectra of this star that clearly revealed an SB2 system with two late O-type components (Fig.~\ref{fig: hd099}). Clear RV variations are seen from one night to the other, suggesting a relatively short period. Both \heb\ and \sid\ lines are definitely present in the secondary spectrum. Disentangling the secondary signature in the \heb\l4542 region is more difficult because of the presence of \nc\ll4534-4545 in the primary spectrum that has a similar strength as the secondary \heb\l4542 line. Reliable separation is achieved on a couple of spectra only. Our best estimate for the observed secondary \heb\l4542 EW  is 0.08\AA, and we used this value to derive the spectral type. Based on our measurements, the spectral type estimates are O9 and O9.7 for the primary and secondary stars respectively, with the O8.5 and O9.5 spectral sub-types well within the respective uncertainties. The \sid\l4089 over \hea\l4144 line ratio, $W''$, indicates that the primary is likely a giant, with the supergiant class within the uncertainties, and that the secondary is likely a main sequence star. Given the object's magnitude ($V=8.09$), a supergiant primary would place the object at 4 to 5~kpc and a class III would place it at about 3~kpc. While HD~100099 is thus potentially a member of a different association \citep{TEH98}, its systemic velocity is compatible with the RVs of the other O stars in IC~2944.

A detailed study of the orbital solution is beyond the scope of this paper. Here we limit ourselves to a preliminary analysis based on the \hea\l5876 line. Our Fourier analysis indicates a most probable period  close to 21.57~d, although we note that the corresponding peak in the periodogram is favoured by the sampling (Fig.~\ref{fig: hd099period}). Best-fit parameters and RV curves are displayed in Table~\ref{tab: hd099os} and in Fig.~\ref{fig: hd099os}, respectively. While the fit is of good quality, our data do not sample the maximum separation epoch, leaving some uncertainties both on the RV curve semi-amplitudes $K_1$ and $K_2$ and on the eccentricity.

% ========================================================================
\begin{figure}
\centering
\includegraphics[width=\columnwidth]{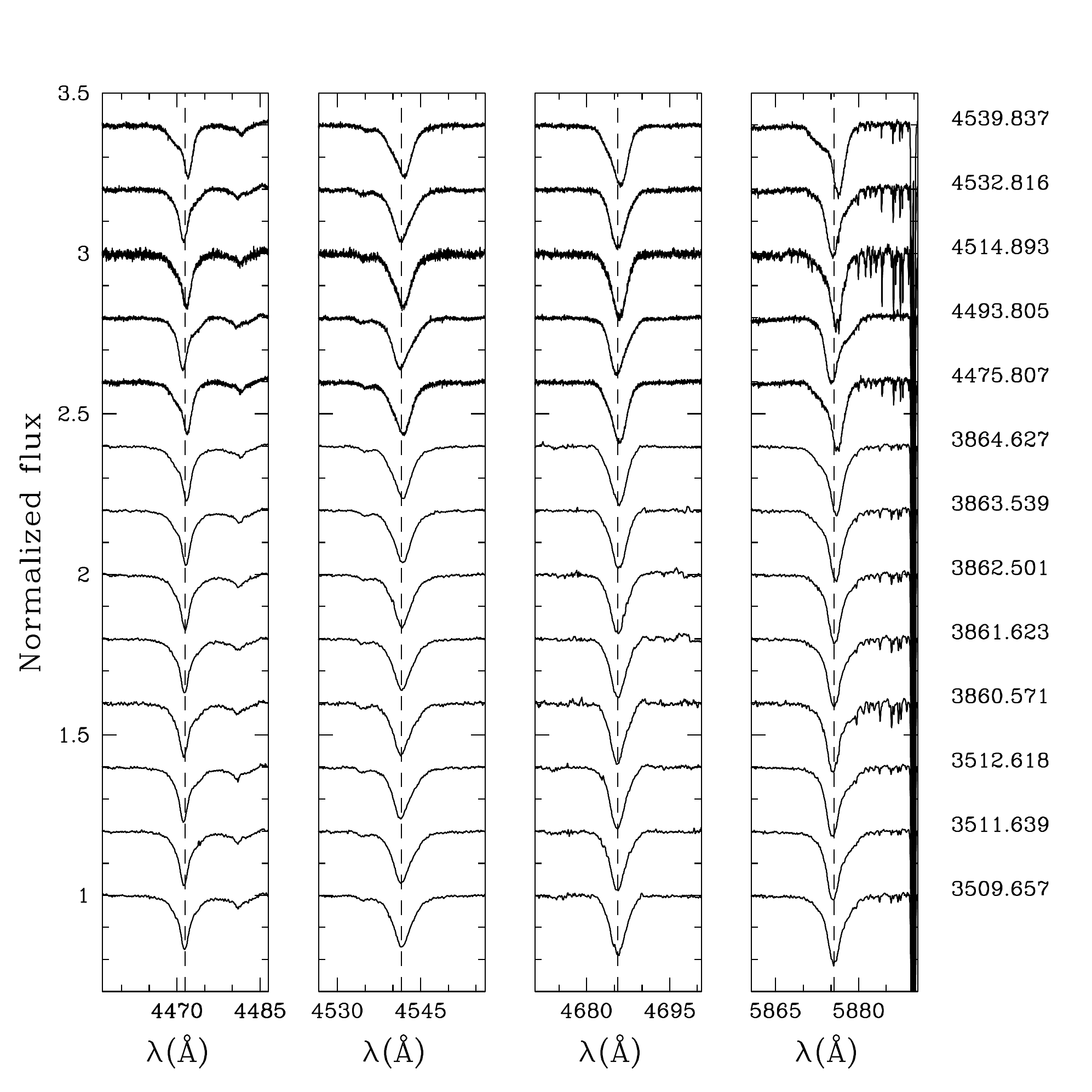}
\caption{{\bf HD~101436:} \hea\l4471, \heb\ll4542, 4686 and \hea\l5876 line profiles at various epochs.}
\label{fig: hd436}
\end{figure}
% ========================================================================

% ========================================================================
\begin{figure}
\centering
\includegraphics[width=\columnwidth]{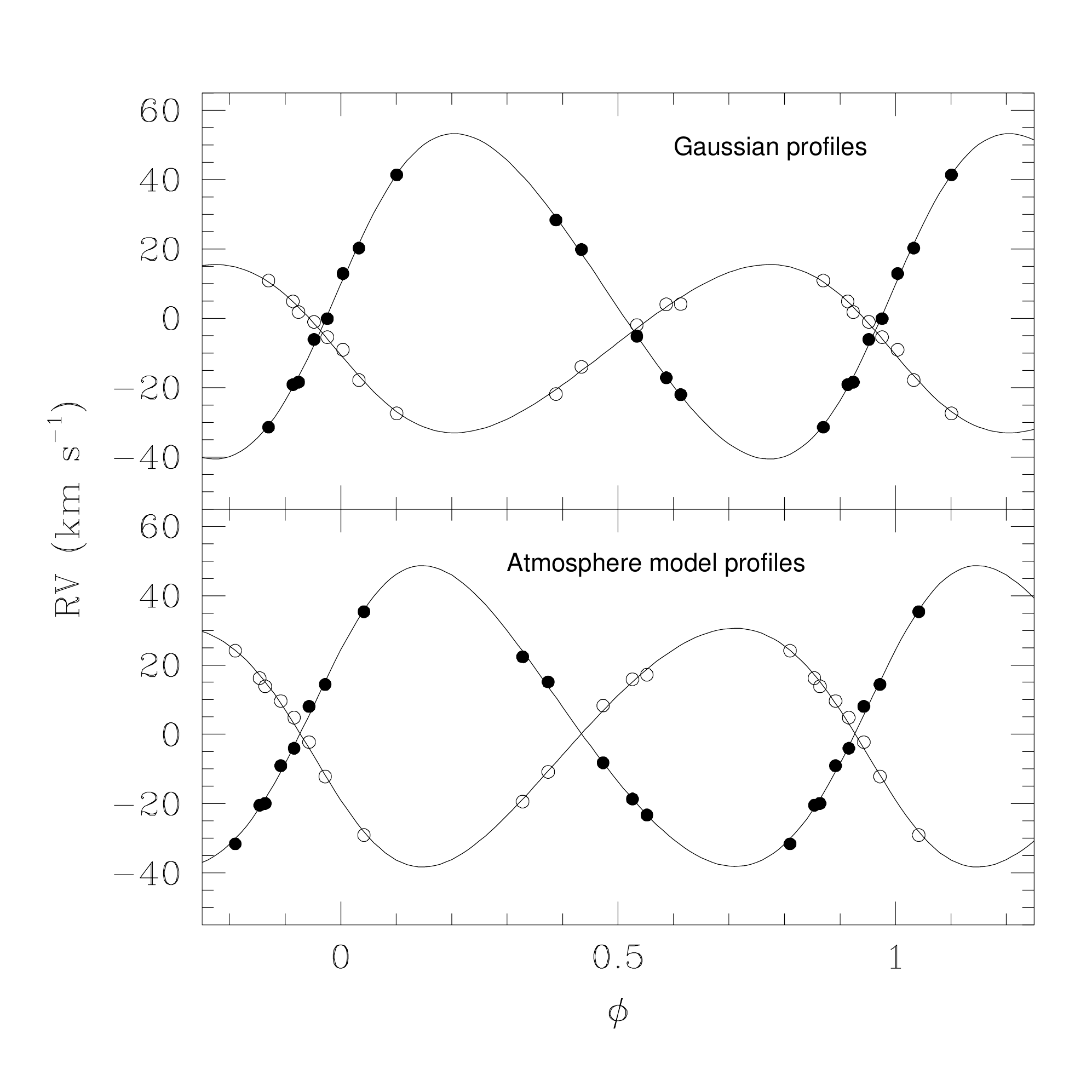}
\caption{{\bf HD~101436:} primary (filled symbols) and secondary (open symbols) RVs together with the best-fit RV curves of Table~\ref{tab: hd436os}.}
\label{fig: hd436os}
\end{figure}
% ========================================================================

\sss{HD~101436}
HD~101436  \citep[O6.5~V, ][]{Wal73} was flagged as a probable RV variable star by \citet{ThW65}, with significant RV changes on a time scale of a week. Two additional RV measurements by \citet{Hum73} and \citet{CLL77} confirmed a $\Delta RV$ amplitude of $\approx$40~\kms. We obtained 13 additional spectra covering time scales from days to years that reveal, for the first time, a shallow secondary signature in the \hea, \heb\ and \sid\ lines (Fig.~\ref{fig: hd436}). Fitted with a single Gaussian, the narrow (primary) component  presents a clear and smooth increase over our 5-day campaign in May 2006, rising by about 40~\kms\ from $-$20 to  $+$20~\kms\ approximately, although slightly depending on the line considered. Yet, RV measurements are hampered by a degeneracy in the determination of the line profiles. Two different hypotheses allow us to fit the data. Either the secondary contributes to the blended profile as a faint and narrow component or it contributes as a broad and shallow one. In both cases however, the values of the Conti's criteria are left unaffected within the uncertainties, with  the primary being an O7 star and the secondary, an O6.5 star with half sub-type uncertainties on both measurements. Because the \hea\l4144 line is very weak, the $W''$ criterion is unsuited and we relied on the $W(\lambda4686)$ measurements to estimate the luminosity class. Depending on the adopted width for the secondary, the \heb\l4686 line is dominated either by the primary or by the secondary star. The dominating component displays a strong absorption, typical of dwarf stars. The other component shows then a measured value of $\log W(\lambda4686)\approx2.2-2.3$.  Because the line is diluted by the companion continuum, this measurement provides only a lower limit on the true strength of the \heb\l4686.  It would correspond to a main sequence star if the light ratio is 1:3 or more. Given the system magnitude, both stars are further expected to be main sequence stars if located in \ic\ and we adopt O7~V+O6.5~V as a tentative spectral classification. 

To double check the results of the RV genetic fitting algorithm, we also developed a global $\chi^2$ minimization approach. We adjusted the profiles of the \hea\ll4026, 4471, 5876 and of the \heb\ll4200, 4542, 4686 lines simultaneously on all the 13 spectra using the following assumptions: (i) the profile of each line is formed by a blend of two Gaussian components representing the contribution of each star, (ii) the RV of each star is identical for all the lines at a given epoch, (iii) the amplitudes and full widths at half maximum (FWHMs) of the Gaussians are identical for a given line and a given stellar component at all epochs.  As a result, this approach definitely favours the second option of a narrow primary and a broad and shallow secondary.

% ========================================================================
\begin{table}
 \centering
 \begin{minipage}{80mm}
  \caption{{\bf HD~101436:} best-fit orbital solution using RVs based on Gaussian profiles (Col.~2) and on atmosphere model profiles (Col.~3). $T$  (in HJD$-$2\,450\,000) is the time of periastron passage. Quoted uncertainties correspond to 1\s\ error-bars.}
\label{tab: hd436os}
  \begin{tabular}{@{}lll@{}}
  \hline
Parameter              & Gaussian profiles             & Atmosphere model profiles   \\
  \hline
$P$ (d)                & $ 37.37   \pm 0.14$ & $ 37.37   \pm 0.14$  \\
$e$                    & $   0.11  \pm 0.01$ & $   0.11  \pm 0.02$  \\
\w\ (\degr)            & $   274   \pm 12  $ & $   296 \pm 11    $  \\
$T$                    & $3190.77  \pm 1.14$ & $3193.04  \pm 1.01$  \\
$M_1/M_2$              & $ 0.52  \pm 0.02$   & $ 0.79  \pm 0.03$    \\ 
$\gamma_1$ (\kms)      & $  6.0 \pm 0.8 $    & $   3.2 \pm 0.8   $ \\
$\gamma_2$ (\kms)      & $ -8.7 \pm 0.5 $    & $   -2.2 \pm 0.6   $ \\
$K_1$ (\kms)           & $ 46.9 \pm 1.4 $    & $   43.4 \pm 1.1   $ \\
$K_2$ (\kms)           & $ 24.3 \pm 0.7 $    & $   34.4 \pm 0.9   $ \\
rms (\kms)             & $     1.3 $ & $1.4$\\
\hline
\end{tabular}
\end{minipage}
\end{table}
% ========================================================================

 As a final step, we replaced the Gaussian profiles by more realistic profiles. The latter were computed using FASTWIND \citep{PUV05} for a range of parameters typical of mid-O stars \citep{MSH05}, then convolved with a rotational profile. Given the high spectral resolution of our data, we neglected the effect of the instrumental profile. Dilution by the companion continuum was included as a free parameter. To increase the realism of the approach, we further required that each line of a given component uses the same dilution factor and the same projected rotation rate. The best fit was obtained by using atmosphere models corresponding to the physical parameters of an O6~V primary with $v \sin i \approx 70$~\kms\ and an O6.5~V secondary with $v \sin i \approx 200$~\kms.

While the overall line profiles are well reproduced in both approaches, the obtained RVs show systematic differences. On average, the RVs associated to the narrow component are 3~\kms\ smaller when using the FASTWIND profiles and the RVs of the broad component are 8~\kms\ larger. This has a significant impact on the respective amplitudes of the RV signal as well as on their ratio (Table \ref{tab: hd436os}). %As shown in Fig~\ref{fig: v1v2}, this implies a significant uncertainty on the mass-ratio of the two components. 

Despite these uncertainties,  the global $\chi^2$ minimization improves the disentangling performances compared to the genetic algorithm described in Sect.~\ref{sect: obs}, allowing us to constrain some of the orbital parameters. For both RV sets, a Fourier analysis indicates two most probable periods of 12.9 and 37.3~d, the second one presenting the highest peak in the periodograms. A preliminary orbital solution analysis also favours the longer period. In both cases, slightly eccentric orbits provide a better fit compared to circular orbits.   All in all, the orbital solutions built on both RV sets yield parameters in good agreement, except for the amplitude of the secondary RV curve and for the mass-ratio.  In particular, the RV measurements based on the more realistic FASTWIND profiles yield a mass-ratio closer to unity, which is in better agreement with our  spectral type estimates. 

Fig.~\ref{fig: hd436os} displays the best-fit RV curves and Table~\ref{tab: hd436os} lists the corresponding parameters. We note that we do not constrain the maximum RV separation, which is a source of uncertainty for the eccentricity and for the semi-amplitude of the RV curves. While we consider that the period is well constrained and that the system has likely a limited or no eccentricity, we regard other parameters as preliminary. 

Under the adopted hypotheses, the component that we have referred to as the secondary appears to be the hotter, most massive component.  Given the low amplitude of the RV curve, the minimum masses are very small, with values slightly below 1~\msol, suggesting an inclination of the order of 20\degr. This implies a deprojected rotational velocity for the fast rotating component of about 600~\kms. One possible channel to produce such a fast rotator is a recent mass transfer from the primary to the secondary that would have increased the secondary mass and spin it up to the critical rotation rate. If this scenario is correct,  HD~101436 could turn out to be a very interesting system to constrain evolutionary scenarios of massive binaries. More data and, in particular a better coverage of the extrema of the RV curves,  and the use of a disentangling algorithm to recover the line profiles independently of any assumption on their shapes are the needed ingredients to verify our preliminary analysis.

% ========================================================================
\begin{figure}
\centering
\includegraphics[width=\columnwidth]{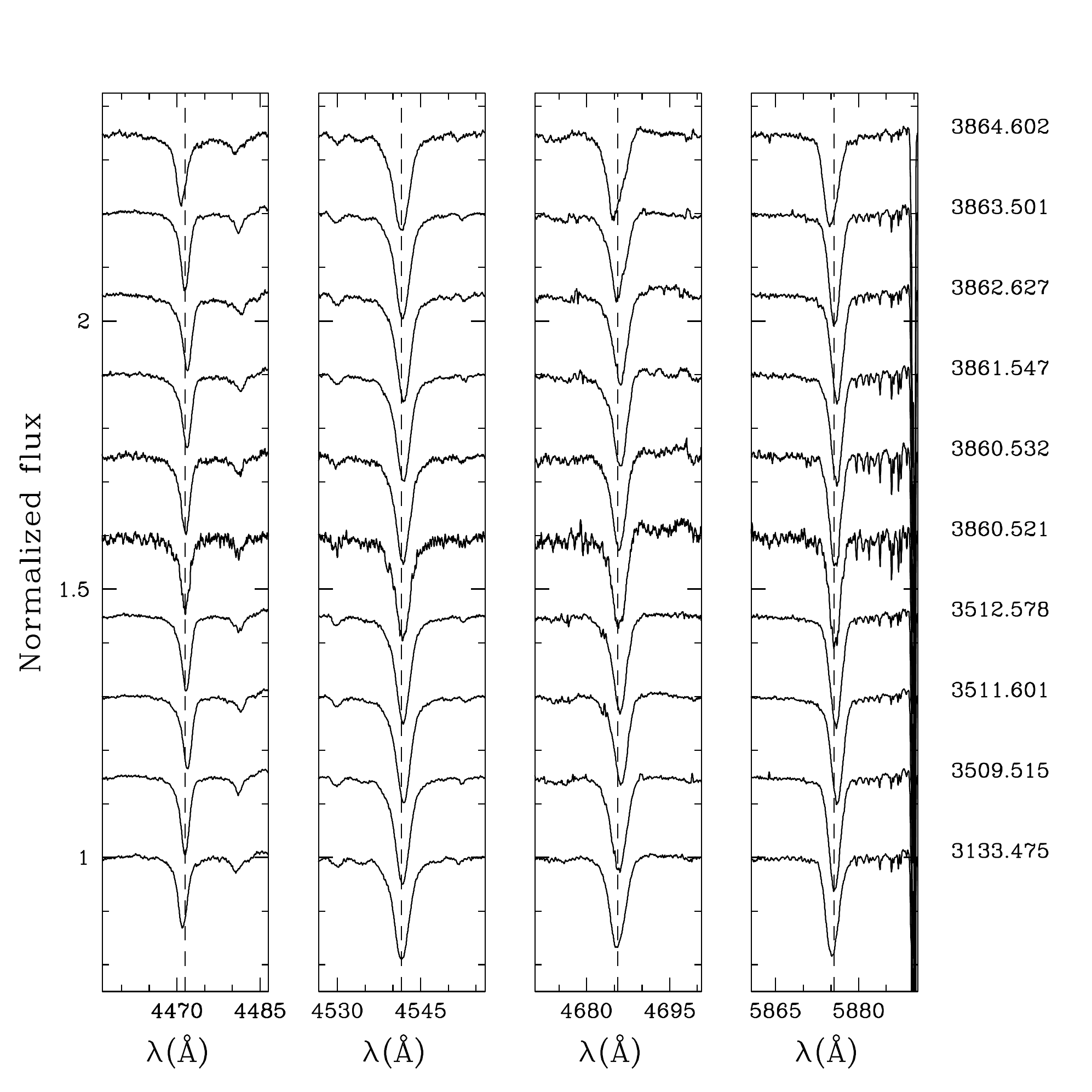}
\caption{{\bf HD~101190:} \hea\l4471, \heb\ll4542, 4686 and \hea\l5876 line profiles at various epochs.}
\label{fig: hd190}
\end{figure}
% ========================================================================

% ========================================================================
\begin{figure}
\centering
\includegraphics[width=\columnwidth]{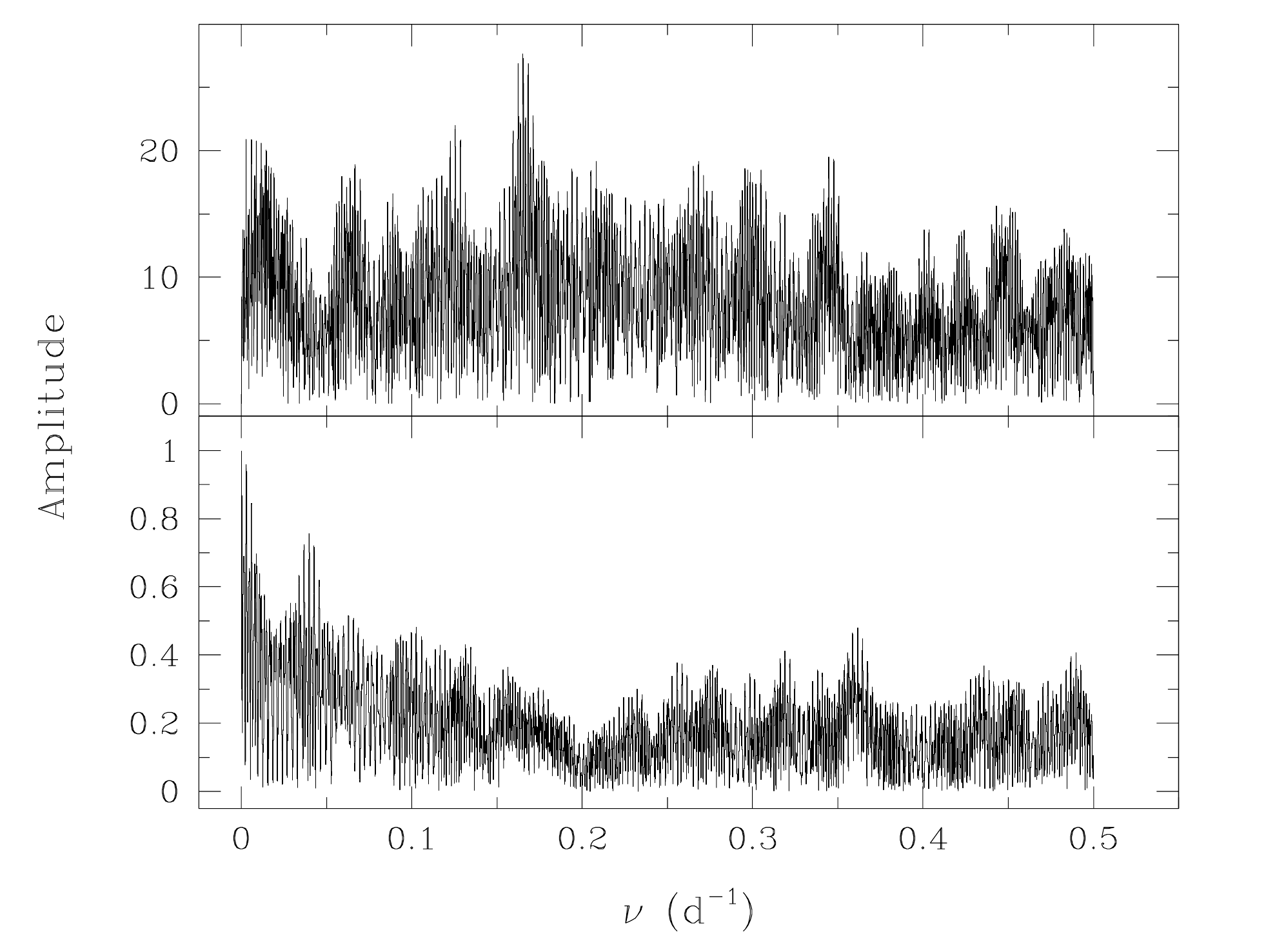}
\caption{{\bf HD~101190:}   periodogram (upper panel) and spectral window (lower panel) computed using the average of the \hea\ll4026, 4471, 4922, 5876 line RVs.}
\label{fig: hd190sp}
\end{figure}
% ========================================================================
% ========================================================================
\begin{table}
 \centering
 \begin{minipage}{80mm}
  \caption{{\bf HD~101190:} best-fit orbital solutions based on the \hea\ and \heb\ data sets. $T$  (in HJD$-$2\,450\,000) 
is the time of periastron passage and is adopted as $\phi=0.0$ in Fig.~\ref{fig: hd190os}. Quoted uncertainties correspond to 1\s\ error-bars.}
\label{tab: hd190os}
  \begin{tabular}{@{}lll@{}}
  \hline
Parameter              & \hea\ lines       & \heb\ lines            \\
  \hline
$P$ (d)                & $ 6.0466\pm0.0026$ & $ 6.0473\pm0.0026$ \\
$e$                    & $   0.29\pm0.04  $ & $   0.36\pm0.07  $ \\
\w\ (\degr)            & $  117.8\pm4.8   $ & $  127.5\pm6.2   $ \\
$T$                    & $2999.17\pm0.09  $ & $2999.13\pm0.11  $ \\
$\gamma_2$ (\kms)      & $   -3.8\pm0.7   $ & $    8.2\pm0.7   $ \\
$K_2$ (\kms)           & $   34.4\pm1.4   $ & $   20.8\pm1.2   $ \\
%$\gamma_2$ (\kms)      & \na                 & \na            \\
%$K_2$ (\kms)           & \na                 & \na            \\
%$q=M_2/M_1$            & \na                 &                 \\
rms (\kms)             & $     2.5        $ & $   2.1         $  \\
\hline
\end{tabular}
\end{minipage}
\end{table}
% ========================================================================

% ========================================================================
\begin{figure}
\centering
\includegraphics[width=\columnwidth]{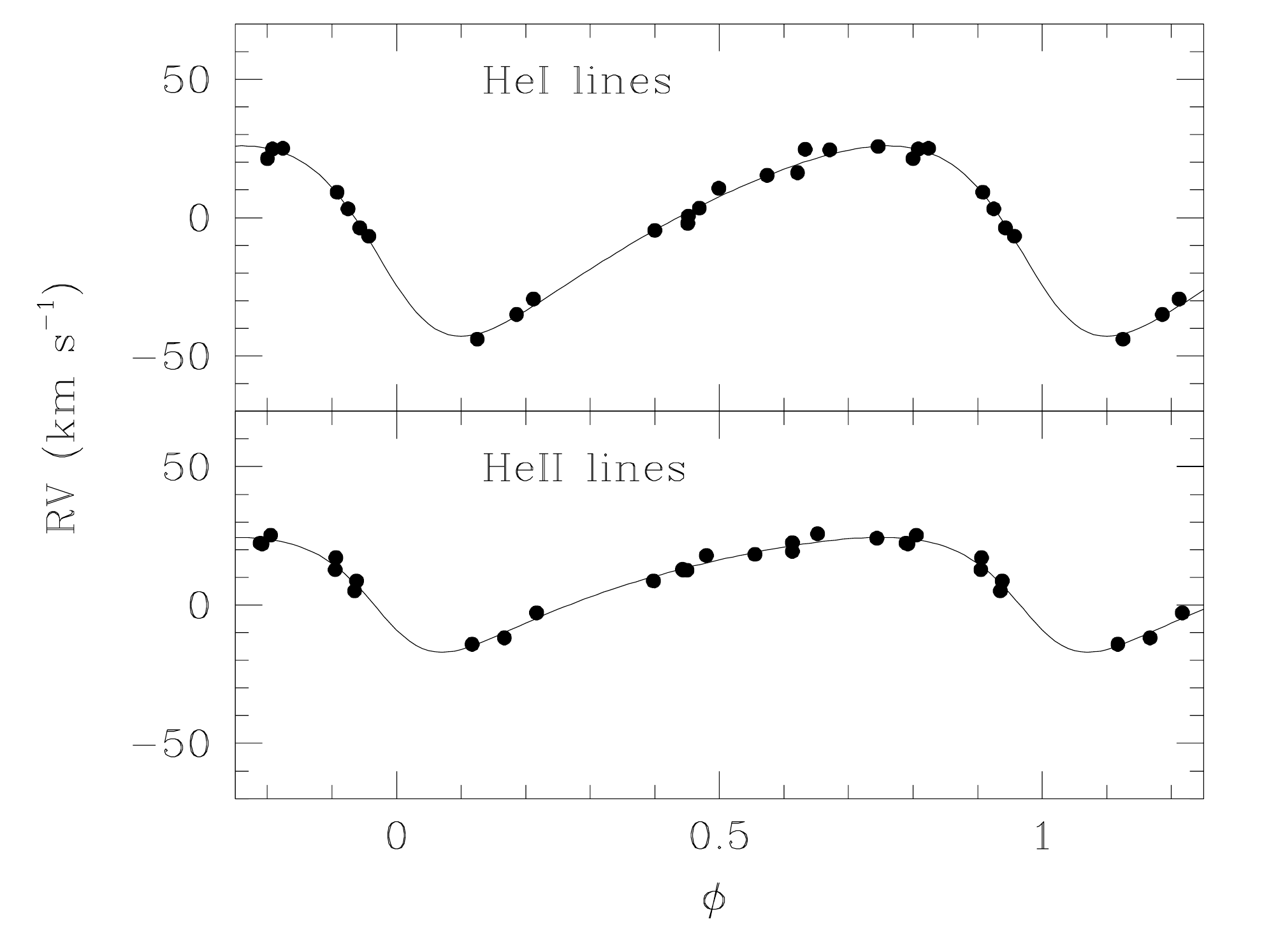}
\caption{{\bf HD~101190:} best-fit RV curve for the \hea\ (upper panel) and \heb\ (lower panel) data sets.}
\label{fig: hd190os}
\end{figure}
% ========================================================================

\sss{HD~101190}
To the North of the cluster, HD~101190 \citep[O6~V((f)), ][]{Wal73} is reported as RV variable star \citep{ArM77}. 
Photometric monitoring by \citet{Bal92} also revealed significant variability with a rms amplitude of 7 millimag although no period has been identified so far. We collected 20 spectra over a 4 year timebase that shows relatively narrow lines, consistent with the rotational velocities deduced by \citet{Pen96} and \citet{HSHP97} (\vsini=89 and 88~\kms respectively). Our data further reveal clear 
variability of all the considered lines with asymmetric profiles seen at several epochs (Fig.~\ref{fig: hd190}).  The \hea\ lines show a larger RV amplitude than the \heb\ lines
but no secondary signature could be reliably disentangled by Gaussian fitting. Our classification criteria assign an O5.5~V((f)) spectral type to the composite spectra.

Inspection of the spectra reveals several interesting features. All the spectral lines in the HD~101190 spectrum display a clearly correlated motion, except the \nc\ll4634-4640 emission lines and the \ne\ll4604-4620 absorption lines that show constant velocity ($\sigma_\mathrm{RV4604, 4620}\approx3$~\kms). On two occasions, the \sid\ll4089, 4116 lines are seen separated into a blue-shifted absorption and a red-shifted emission, the latter emissions displaying RVs consistent with those of the \nc\ and  \ne\ lines. The presence of the \ne\ll4604-4620 absorptions and of the \sid\ll4089-4116 emissions is not compatible with the O6~V((f)) type reported by \citet{Wal73}, nor is it compatible with the average of the \hea\l4471 over \heb\l4542 ratio measured from our data. This suggests that the HD~101190 spectrum has two components: a very early O-type star that shows no RV variations and that is responsible of the \ne\ absorptions and \sid\ emissions, and a later O-type star which carries the RV signal.

Because we cannot detect the \nd\l4058 line and because the \ne\ and \sid\ lines are relatively faint, we assign the hotter star an O4~V((f+)) type, although we considered the O3.5 sub-type within the uncertainties. Given the expected visible flux ratio between an O4~V star and a less massive O7 component \citep[$\approx2.75$,][]{MSH05}, one finds that an O7-O7.5~V star, with $W'\approx1$, can reproduce the observed $W'$ ratio of the composite spectrum. While our data do not allow us to prove that the O4 and O7 components are physically bound, we note that the chance alignment of two such stars is very small.

In an attempt to constrain the orbital properties of the system, we performed a Fourier analysis of the RV series associated with the individual \hea\ and \heb\ lines. All the data sets consistently yield an orbital period of 6.047~d, which is indeed the dominant alias in the periodogram (Fig.~\ref{fig: hd190sp}). We computed SB1 orbital solutions both using the RVs from the individual lines and by combining separately the \hea\ll4026, 4471, 4922, 5876  and the \heb\ll4200, 4542, 4686 RV measurements. Again we obtained very similar results. Table~\ref{tab: hd190os} and Fig.~\ref{fig: hd190os} give the best fit orbital solution for the \hea\ and \heb\ data sets and show the corresponding RV curves. While we consider that the orbital period is reliable, we note that the semi-amplitude of the RV curve $K$, is likely a lower limit as it is affected by the blend with the O4 signature. 

Finally, assuming that the secondary is not a very late O-star, an hypothesis supported by the RV signatures in the \heb\ lines, the binary mass-ratio $M_2/M_1$ is likely larger than 0.3. This would imply a lower limit on the primary RV curve semi-amplitude: $K_1>10$~\kms, so that one would have expected to see some variations of the primary lines. While this argues against  a direct bound between the O4 and  the O7 star, we note that a mass ratio a factor of two lower than assumed here would solve this apparent discrepancy. Alternatively, the system may be triple, with a single hot star and an SB1 cooler star.  More data are definitely needed to solve this issue.

% ========================================================================
\begin{figure}
\centering
\includegraphics[width=\columnwidth]{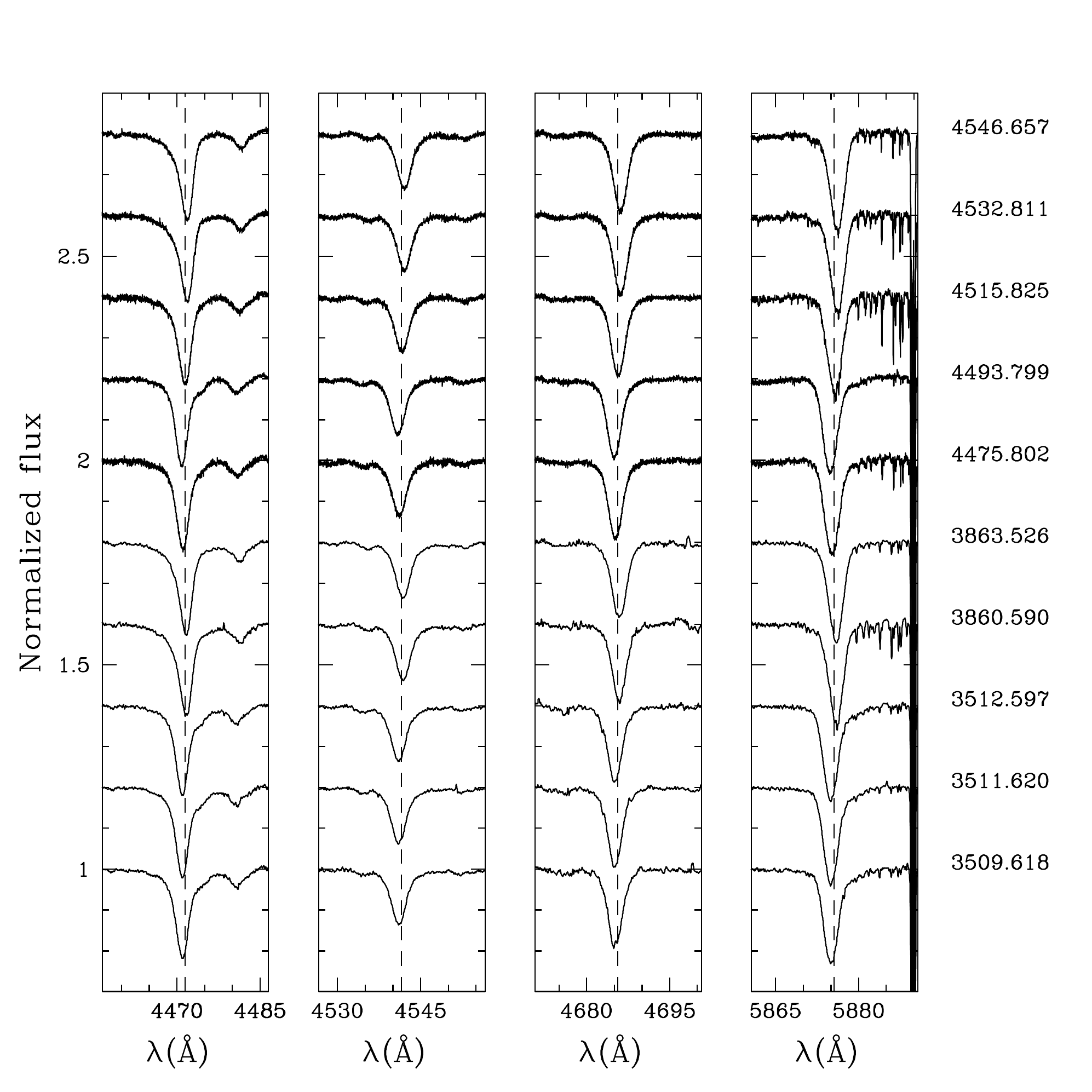}
\caption{{\bf HD~101413:} \hea\l4471, \heb\ll4542, 4686 and \hea\l5876 line profiles at various epochs.}
\label{fig: hd413}
\end{figure}
% ========================================================================

\sss{HD~101413}
HD~101413 \citep[O8~V, ][]{Wal73} is the visual companion of HD~101436, with a separation of 27.8\arcsec\ \citep{MGH98}. \citet{ThW65} first reported large RV variations ($\Delta RV\sim100$~\kms) based on 3 spectra collected over a year. IUE and Fuse rotational velocities were measured in the range 90-102~\kms \citep{Pen96, HSHP97, PeG09} and probably relate to the primary or to the composite spectrum. We obtained three \feros\ spectra in May 2005, two  in May 2006 and a set of 5 \uves\ spectra spread over 3 months early 2008. The 2005 and some of the 2008 observations revealed the signature of the secondary component for the first time, although only in the \hea\ lines (Fig.~\ref{fig: hd413}).  A limited RV difference is seen on a day-to-day basis, but the primary and secondary smoothly varied from Jan to Mar 2008, suggesting a period of 3 to 6 months. We note that we never observed such an extreme primary RV as the one reported by \citet{ThW65} in Apr 1957 ($RV_\mathrm{prim}=-81$~\kms). Using the few well separated spectra, we confirm the primary O8~V classification. The companion spectrum shows neither \heb\ nor \sid\ lines and we assign it a B2-3~V classification. While we are missing a complete orbital coverage, empirical determination of the mass-ratio can still be obtained. Based on the \hea\ lines, we obtained $M_2/M_1=0.17\pm0.01$, which rather suggests a mid- to late-B type for the secondary.

%%%%%%%%%%%%%%%%%%%%%%%%%%%%%%%%%%%%%%%%%%%%%%%%%%%%%%%%%%%%%%%%%%%%%
\subsection{SB1 systems}\label{sect: SB1}

% ========================================================================
\begin{figure}
\centering
\includegraphics[width=\columnwidth]{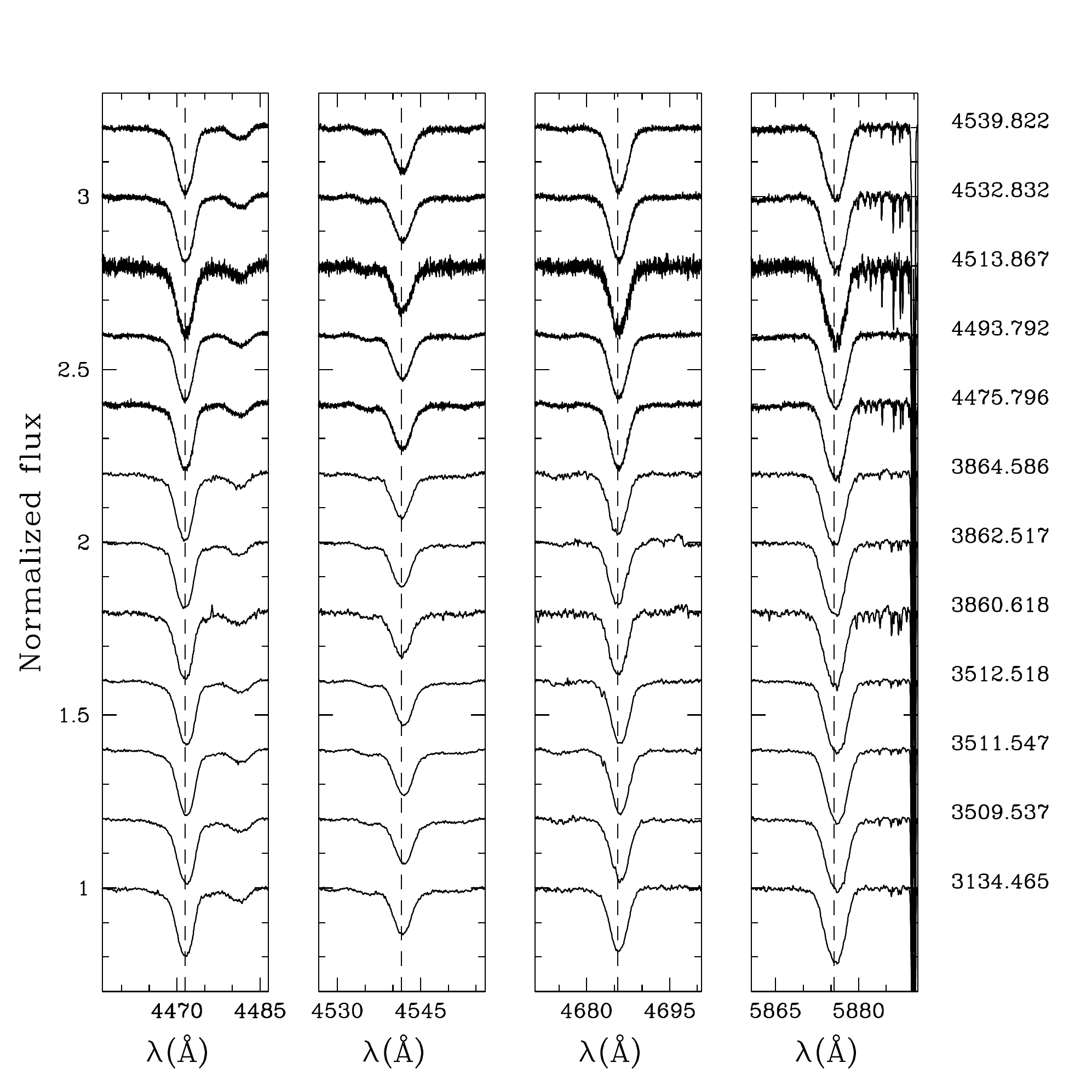}
\caption{{\bf HD~101191:} \hea\l4471, \heb\ll4542, 4686 and \hea\l5876 line profiles at various epochs.}
\label{fig: hd191}
\end{figure}
% ========================================================================

\sss{HD~101191}
HD~101191 \citep[O8~V((n)),][]{Wal73} is located at 1.4\arcmin\ SW of HD~101205, in the core of the cluster and has been scarcely studied so far. We obtained 12 high-resolution spectra that reveal a clear SB1, long period signature (Fig.~\ref{fig: hd191}). Our different epochs indeed reveal RV variations with a peak-to-peak amplitude of $\approx$30~\kms\ for the \heb\ lines and  of $\approx$20~\kms\ for the \hea\ lines. No significant night to night variability is observed, nor within the 3-month coverage of our \uves\ data. While we cannot rule out an alias at 20~d, such a period value is unlikely because the 3- and 5-day \feros\ campaigns in May 2005 and May 2006 reveal no RV variation at all. This rather suggests a long orbital period, of several times the 3-month timebase of our \uves\ campaign. From our data, we can also place a relatively firm upper limit on the orbital period of about 4~yr. As the measured  RVs display a relatively limited amplitude, we might have missed the epoch of maximum separation, keeping open the possibility that one can still observe the signatures of the two components. We derived an O8~V/III type for the composite spectrum. Because the \heb\l4686 line is strongly seen in absorption, we adopt the O8~V spectral sub-type as our final classification of the HD~101191 composite spectrum.

% ========================================================================
\begin{figure}
\centering
\includegraphics[width=\columnwidth]{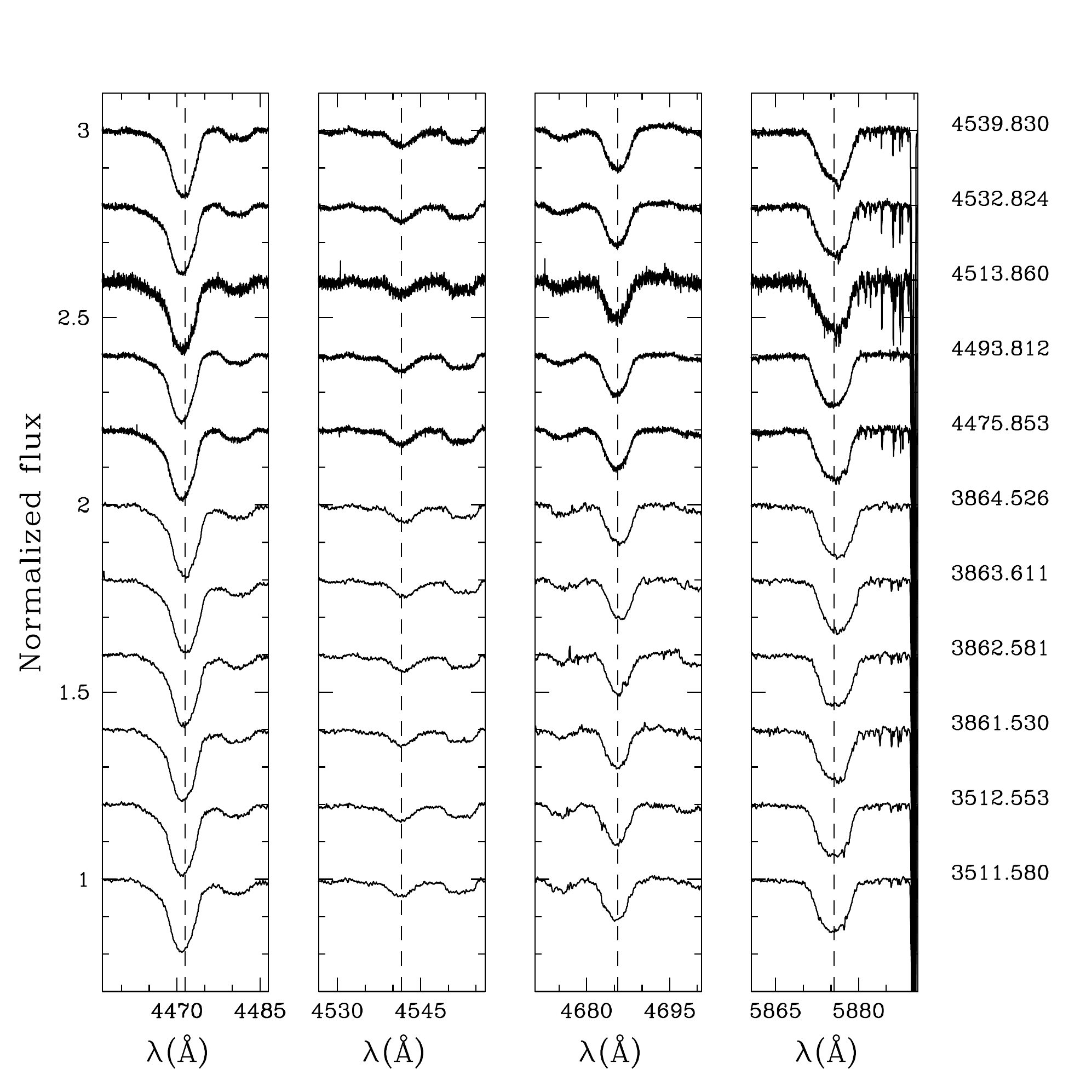}
\caption{{\bf HD~308813:} \hea\l4471, \heb\ll4542, 4686 and \hea\l5876 line profiles at various epochs.}
\label{fig: hd813}
\end{figure}
% ========================================================================

\sss{HD~308813}
At 4.2\arcmin\  NW of HD~101205, HD~308813 was classified O9.5~V by \citet{Sch70} and B0~V by \citet{ArM77}. \citet{ThW65} observed it three times over a 2 yr time-scale and reported $\Delta RV\approx 28$~\kms. Based on discrepant \heb\ RVs compared to the \hea\ ones, they proposed that HD~308813 is an SB2 candidate.  \citet{HuG06a} also collected three observations over 4 days  and reported RVs from 21.9 to $-17.6$~\kms, confirming its SB1 nature. They further measured a projected rotational velocity \vsini\ of 196~\kms. We collected 11 spectra of HD~308813, covering time scales from days to years.
HD~308813 presents broad, somewhat variable lines (Fig.~\ref{fig: hd813}). We measured RVs using both Gaussian profile and purely rotationally broadened profiles and obtained very comparable results. Clear RV shifts (peak-to-peak difference over 40~\kms) are seen, although these are not happening as quickly as reported by \citet{HuG06a}. All the lines follow the same orbital motion, so that we cannot confirm the SB2 nature of the object. We however note that some lines display systematic shifts in their RVs compared to the average. Given our data set, we cannot distinguish between a very-short period system ($P<2$~d) and a longer one ($P>20$~d). With a clear \heb\l4542 line, HD~308813 is not a B star but a late O star. We thus confirm the O9.5~V classification of \citet{Sch70}.

%%%%%%%%%%%%%%%%%%%%%%%%%%%%%%%%%%%%%%%%%%%%%%%%%%%%%%%%%%%%%%%%%%%%%
\subsection{Presumably single stars } \label{sect: sgl}

% ========================================================================
\begin{figure}
\centering
\includegraphics[width=\columnwidth]{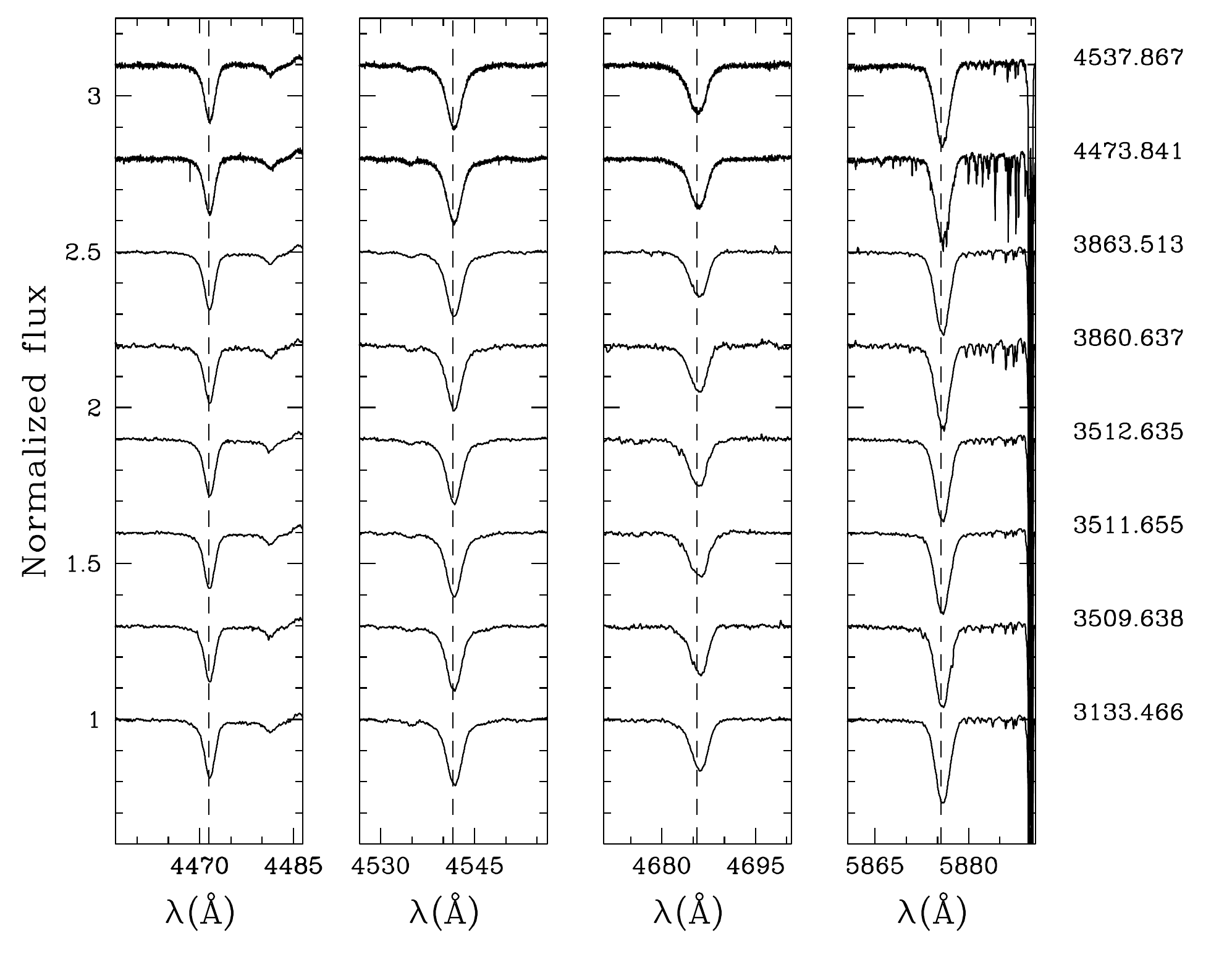}
\caption{{\bf HD~101298:} \hea\l4471, \heb\ll4542, 4686 and \hea\l5876 line profiles at various epochs.}
\label{fig: hd298}
\end{figure}
% ========================================================================

\sss{HD~101298}
The object was observed spectroscopically by \citet{ThW65}, \citet{CLL77} and \citet{ArM77}. The latter reported possible RV variations within their  four measurements. A 6-night photometric monitoring \citep{Bal92} however showed a constant luminosity. Similarly, no significant RV difference is observed in between the eight epochs of our campaign, which again suggests that the star is probably single (Fig.~\ref{fig: hd298}). Spectral type estimates point toward an O6-6.5 star. The $W''$ criterion is unsuited for such an early star as the \hea\l4144 line has become very weak. With $\log W(\lambda4686)\approx2.7$, the \heb\l4686 line suggests a giant classification. Being slightly variable while the other lines remain constant, the \heb\l4686 line  further suggests the presence of a significant wind, another hint of a slightly more evolved object. Following \citet{Wal73}, we adopt an O6~III((f)) classification. Note however that a class V would be in better agreement with the overall distance to the cluster (see Sect.~\ref{ssect: dist}). 

% ========================================================================
\begin{figure}
\centering
\includegraphics[width=\columnwidth]{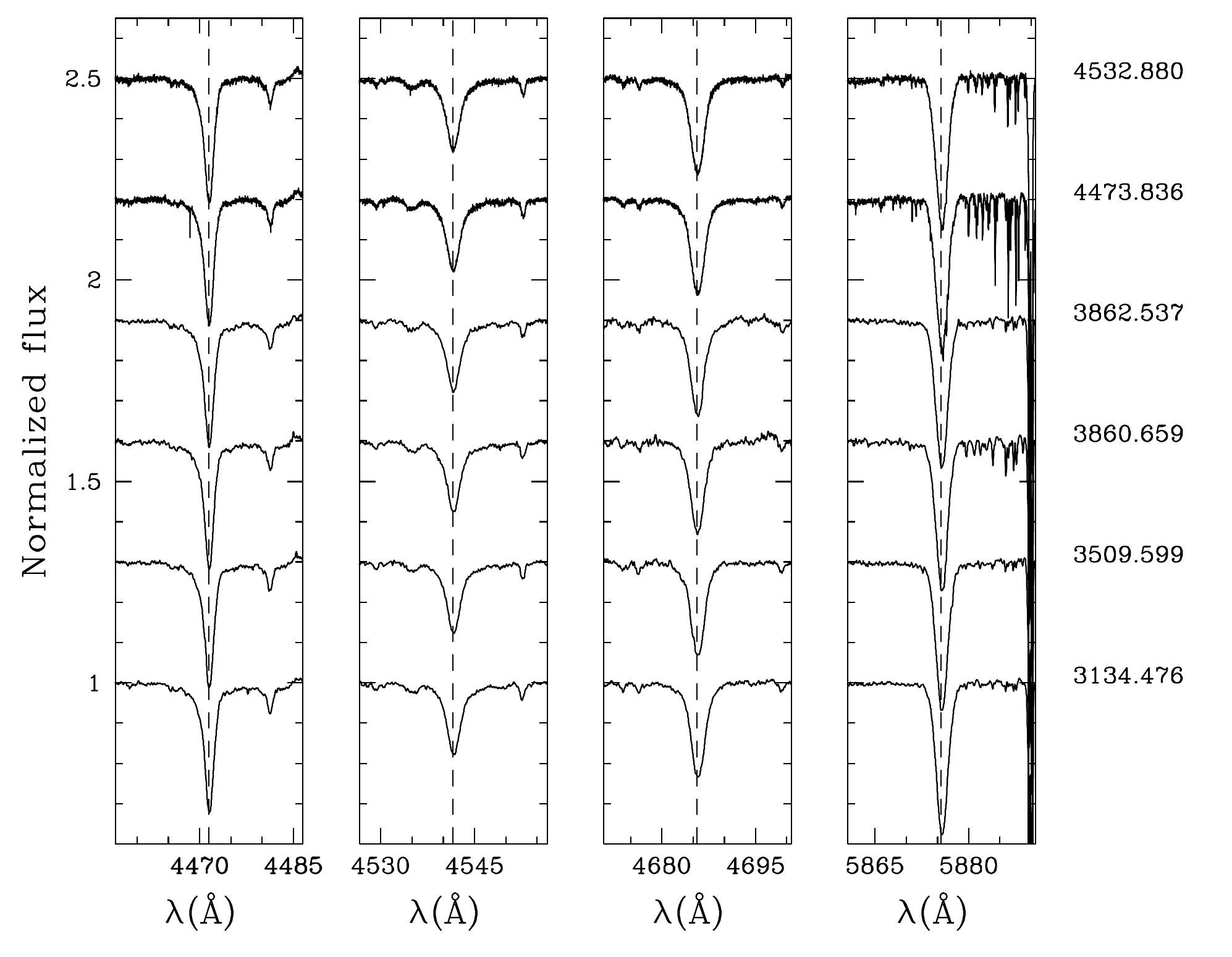}
\caption{{\bf HD~101223:} \hea\l4471, \heb\ll4542, 4686 and \hea\l5876 line profiles at various epochs.}
\label{fig: hd223}
\end{figure}
% ========================================================================

\sss{HD~101223}
\citet{ArM77} reported discrepant velocity measurements (ranging from $-$29 to $+$4~\kms) and asymmetrical line profile, suspecting HD~101223 to be a spectroscopic binary. Observed at six epochs over five years (Fig.~\ref{fig: hd223}), the HD~101223 spectrum remains constant with a 1\s\ RV dispersion of  the order of 1~\kms. The classification criteria indicate an O8~V/III star. Because  the \heb\l4686 line is strongly seen in absorption, we prefer the O8~V classification and we consider the star as single.

% ========================================================================
\begin{figure}
\centering
\includegraphics[width=\columnwidth]{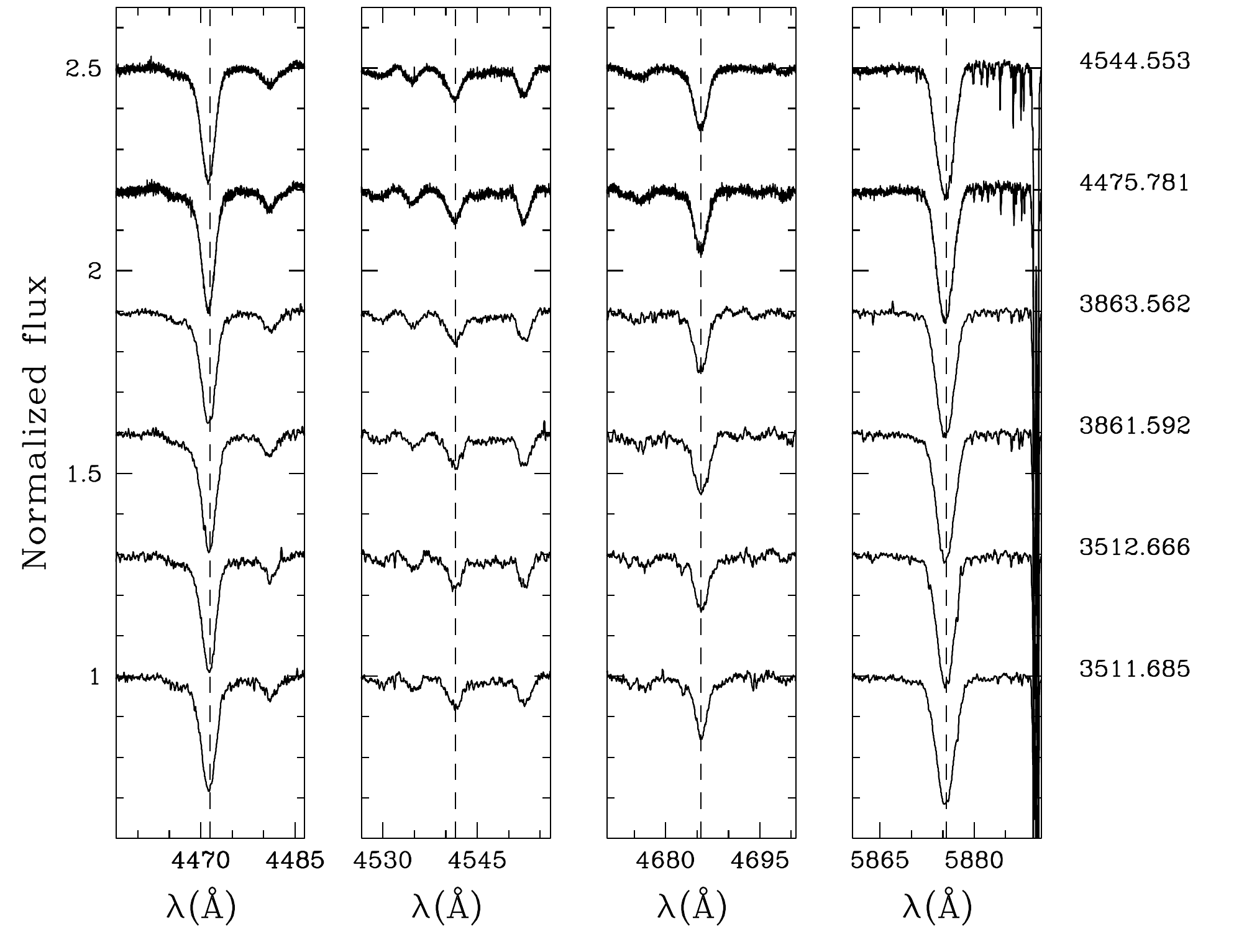}
\caption{{\bf CPD$-$62\degr2198:} \hea\l4471, \heb\ll4542, 4686 and \hea\l5876 line profiles at various epochs.}
\label{fig: cpd2198}
\end{figure}
% ========================================================================

\sss{CPD$-$62\degr2198}
At about 30\arcmin\ NW of the cluster center, \cpd2198 presents a significantly different RV compared to  other single stars (from about $-$6 to $-$12~\kms\ depending on the considered lines). Yet, we could not detect any significant RV variations within our 6 observations (Fig.~\ref{fig: cpd2198}) nor by comparison with older data by \citet{ReK97} and \citet{HuG06a}. 
Classified as O9.5~Ib by \citet{ArM77}, our \hea\l4471 over \heb\l4542 ratio indicates an O9.7 star, with the O9.5 sub-type at 1\s. 
The $W(\lambda4089)/W(\lambda4144)$ ratio indicates a class III star while the $\log W'''$ criterion rather points to a class I star with class III within 1\s.
\citeauthor{HuG06a} (\citeyear{HuG06b}) obtained $T_\mathrm{eff}=29132\pm375$~K and $\log g=3.556\pm0.039$ for CPD$-$62\degr2198, in good agreement with values expected for an O9.7~III star. We consider CPD$-$62\degr2198 to be likely single and we finally adopt the O9.7~III classification. 
% SNR of FEROS is 100-150 while UVES is 150-160 @4800Angstr

% ========================================================================
\begin{figure*}
\centering
\includegraphics[width=1.5\columnwidth]{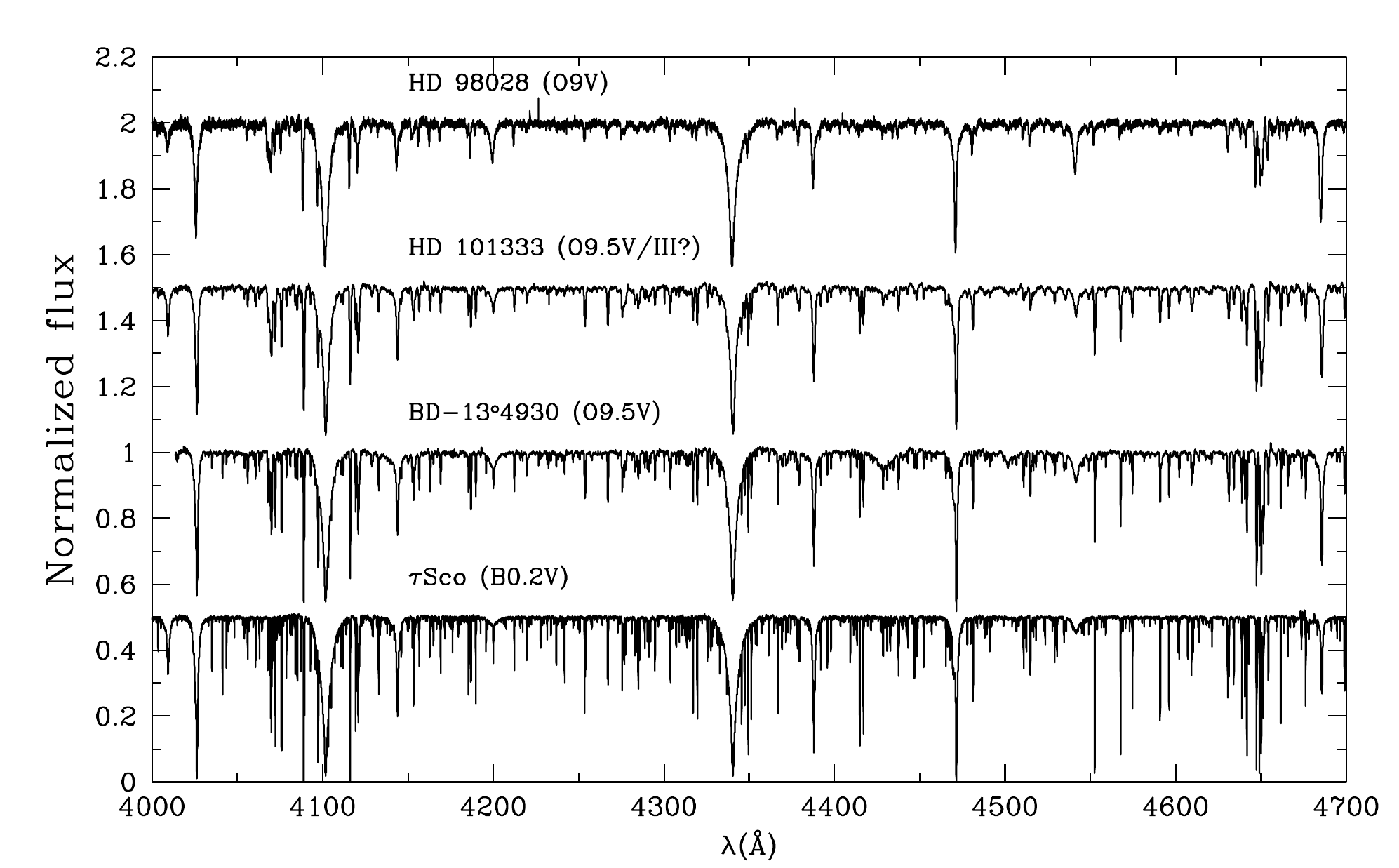}
\caption{{\bf HD~101333}: 4000-4700~\AA\ spectral range compared to the spectroscopic standards $\tau$~Sco and HD~98028 and to BD$-$13\degr4930.}
\label{fig: hd333_compar}
\end{figure*}
% ========================================================================

\sss{HD 101333}
Classified B0~III, HD~101333 is a slow rotator \citep[$v \sin i=39$~\kms, ][]{HuG06a} in the Southern part of the cluster. \citet{ThW65} and \citet{HuG06a} both reported a set of three RV measurements that show no internal discrepancy but these three sets differ by 15~\kms\ from each other. We obtained three \feros\ spectra in May 2006 and another two \uves\ spectra in January and March 2008.  Several lines (e.g.\ \hea\ll4144, 4388) display a clear Lorentz profile while others have a well pronounced Gaussian shape. The UVES data are about 5~\kms\ larger than the \feros\ measurements. Given the slow rotation, this is definitely significant although we cannot decide between a low-amplitude photospheric effect and a large mass-ratio and/or low inclination binary scenario. The obtained spectral classification is somewhat dependent on the assumed line profile, but points to a late-type O star.  Comparing HD~101333  with the spectral standards $\tau$~Sco (B0.2~V) and HD~98028 (O9~V) from \citet{WF90} and to BD$-$13\degr4930 (O9.5 V) from \citetalias{SGE09} (Fig.~\ref{fig: hd333_compar}), we finally adopt an O9.5~V classification.

% ========================================================================
\begin{figure*}
\centering
\includegraphics[width=\columnwidth]{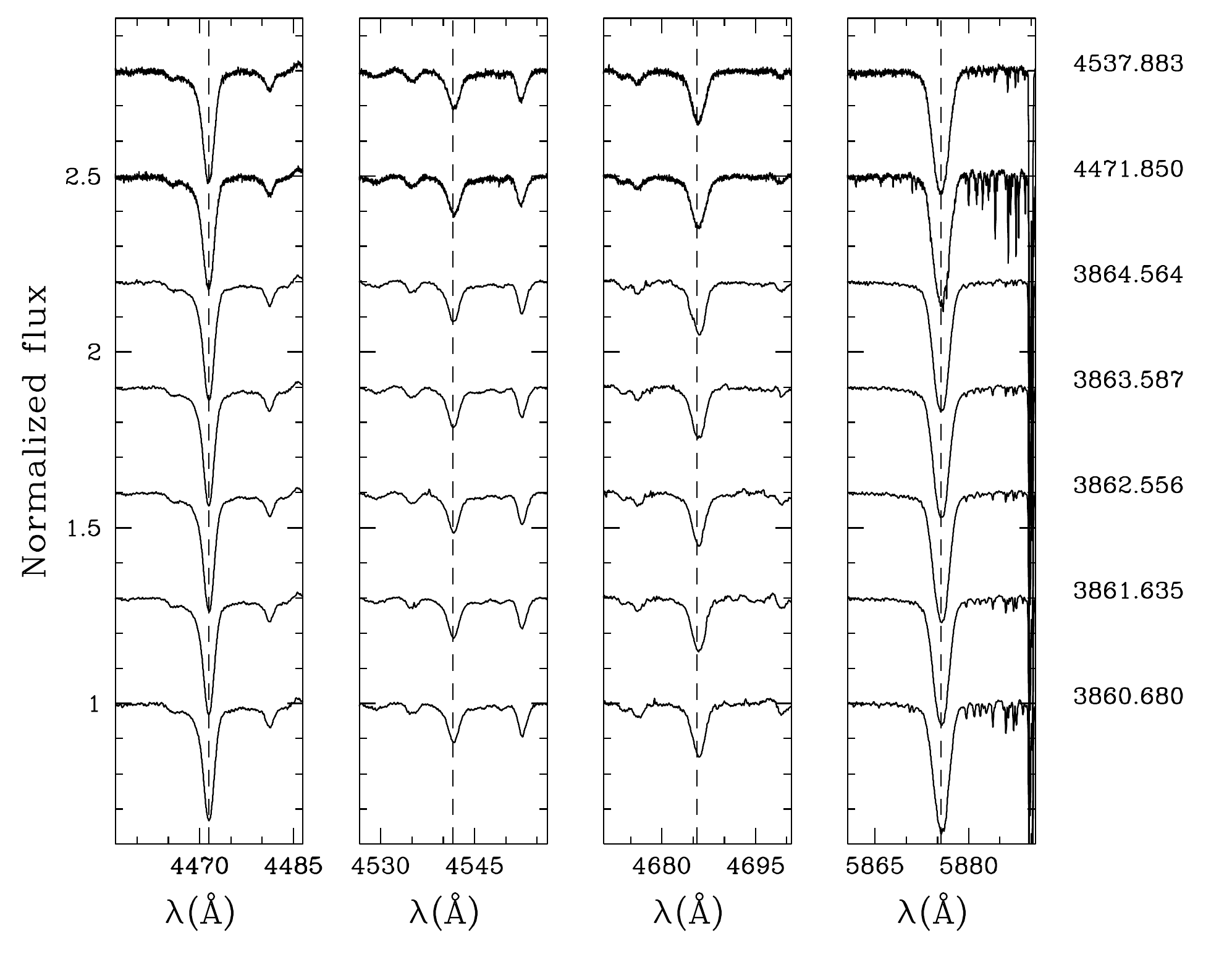}
\includegraphics[width=\columnwidth]{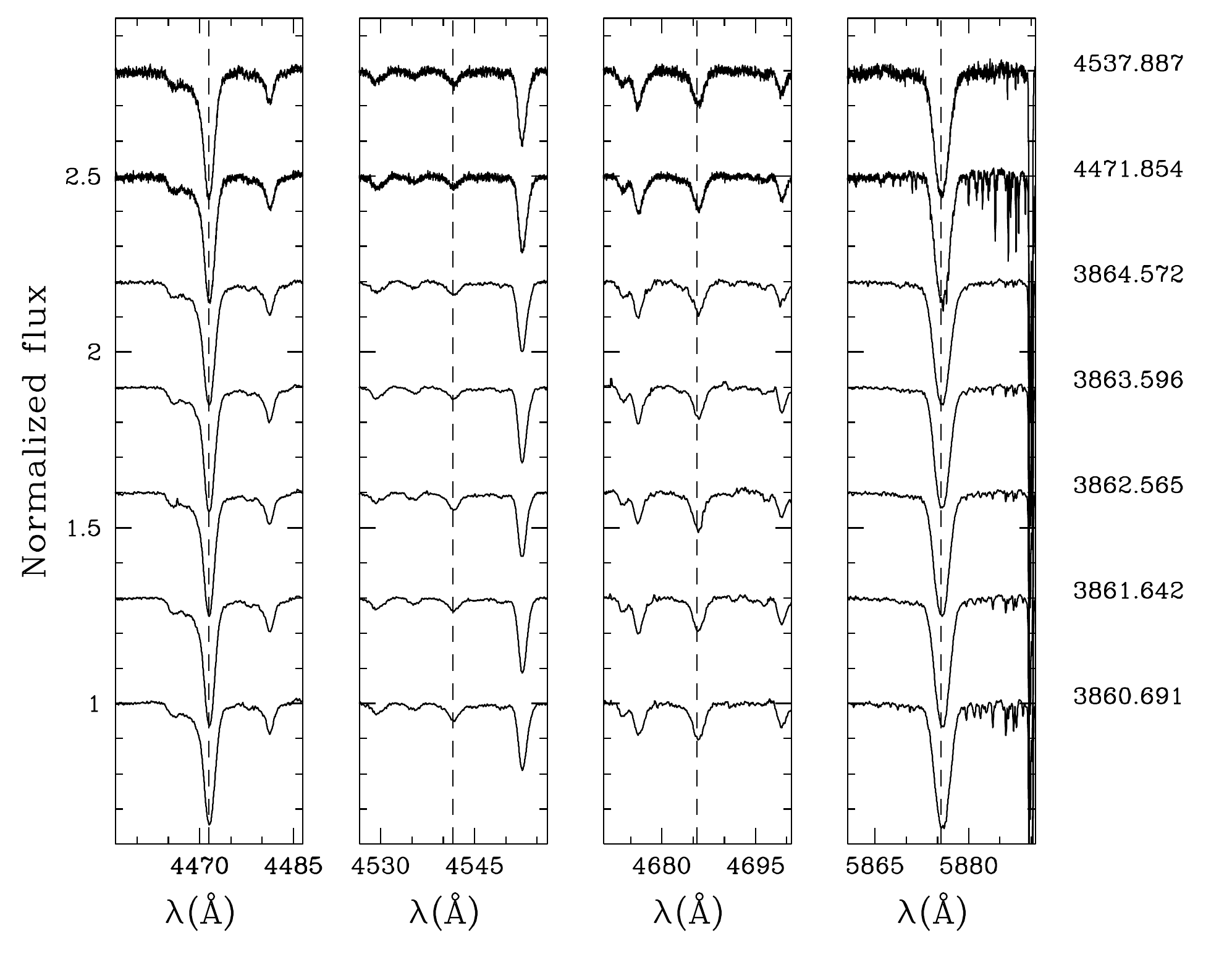}
\caption{{\bf HD~101545A} (left) and {\bf HD~101545B} (right): \hea\l4471, \heb\ll4542, 4686 and \hea\l5876 line profiles at various epochs.}
\label{fig: hd545}
\end{figure*}
% ====================================================

\sss{HD 101545A \& B  }
HD 101545 is a visual binary the components of which are separated by a few arcsecs, the component A being to the NE of component B. \citet{MGH98} reported a separation of 2.57\arcsec\ and respective $V$ mag of 6.9 and 7.4. Both components are late O-type stars (Fig.~\ref{fig: hd545}). None of them display any significant RV variation in their respective time series (7 spectra over 5~yr for each object of the pair). The best spectral type estimate yields O9.5~III/I and O9.7III/I for the A and B components. Given the stars' brightness, we finally consider both stars to be giants and not supergiants. 

%Reported as a Blue Sraggler by Ahumada \& Lapasset A\&A 463, 789-797 (2007) , Based on ThW65 photoelectric measurements, reported as BOIII

\subsection{Other stars in IC~2944}
\sss{HD~308804}
Reported as O9.5~V by \citet{ArM77}, we observed HD~308804 twice with a 2~d separation. The HD~308804 spectrum contains neither \heb\ nor \sid\ lines and it is thus not the spectrum of an O star. While \citet{ArM77} mentioned double or asymmetric lines, our two spectra are single-lined. Yet, they display a RV difference of $\sim10$~\kms, although the first spectrum has a much lower SNR compared to the second one. \citet{HuG06a} also observed this star three times over a four-night period and reported no RV change. Their measured RV is however 10 and 20~\kms\ more negative than our two measurements, providing further support to the binary hypothesis. By comparison with the atlas of \citet{WF90}, we adopt a B3~V type for the composite spectrum of HD~308804.

%%%%%%%%%%%%%%%%%%%%%%%%%%%%%%%%%%%%%%%%%%%%%%%%%%%%%%%%%%%%%%%%%%%%%
%%%%%%%%%%%%%%%%%%%%%%%%%%%%%%%%%%%%%%%%%%%%%%%%%%%%%%%%%%%%%%%%%%%%%

% ========================================================================
\begin{table}
\centering
\begin{minipage}{\columnwidth}
\caption{Binary detection probability for the time sampling associated with different objects (Col.~1) and for various period ranges (Cols. 2 to 5).}
\label{tab: mc_indiv}
\begin{tabular}{ccccc}
\hline
Time sampling & Short   & Intermed. & Long & All \\
          & [2-10d] & [10-365d] & [365-3000d] &[2-3000d] \\   
\hline
HD~100099     &  0.995 & 0.920 & 0.631 & 0.902 \\
HD~101131     &  0.995 & 0.970 & 0.813 & 0.954 \\
HD~101190     &  0.996 & 0.944 & 0.730 & 0.930 \\
HD~101191     &  0.992 & 0.906 & 0.614 & 0.894 \\
HD~101205     &  0.995 & 0.893 & 0.560 & 0.884 \\
HD~101223     &  0.988 & 0.864 & 0.607 & 0.877 \\
HD~101298     &  0.994 & 0.905 & 0.688 & 0.910 \\
HD~101333     &  0.985 & 0.759 & 0.324 & 0.789 \\
HD~101413     &  0.992 & 0.900 & 0.530 & 0.876 \\
HD~101436     &  0.995 & 0.915 & 0.568 & 0.891 \\
HD~101545A    &  0.991 & 0.783 & 0.376 & 0.810 \\
HD~101545B    &  0.992 & 0.785 & 0.341 & 0.805 \\
HD~308813     &  0.992 & 0.886 & 0.463 & 0.859 \\
\cpd2198     &  0.987 & 0.839 & 0.511 & 0.851 \\ 
\hline
\end{tabular}
\end{minipage}
\end{table}
% ========================================================================
%          11       0.121
%          12       0.268
%          13      0.3528
%          14      0.2082
%Midas 001> com .3528+.2082+.268 ==> 83 % conf interval :f~ [0.86-1.0] (or .93+/-0.07)
%  0.8290000    
%Midas 002> com .3528+.2082+.268+.121 ==> 95 % conf interval :f~ [0.79-1.0]
%  0.9500000
%Midas 002> com 13/14 
%  0.9285714
%Midas 003> com 12/14
%  0.8571429
%Midas 004> com 11/14
%  0.7857143
%
% ========================================================================
\begin{figure}
\centering
\includegraphics[width=\columnwidth]{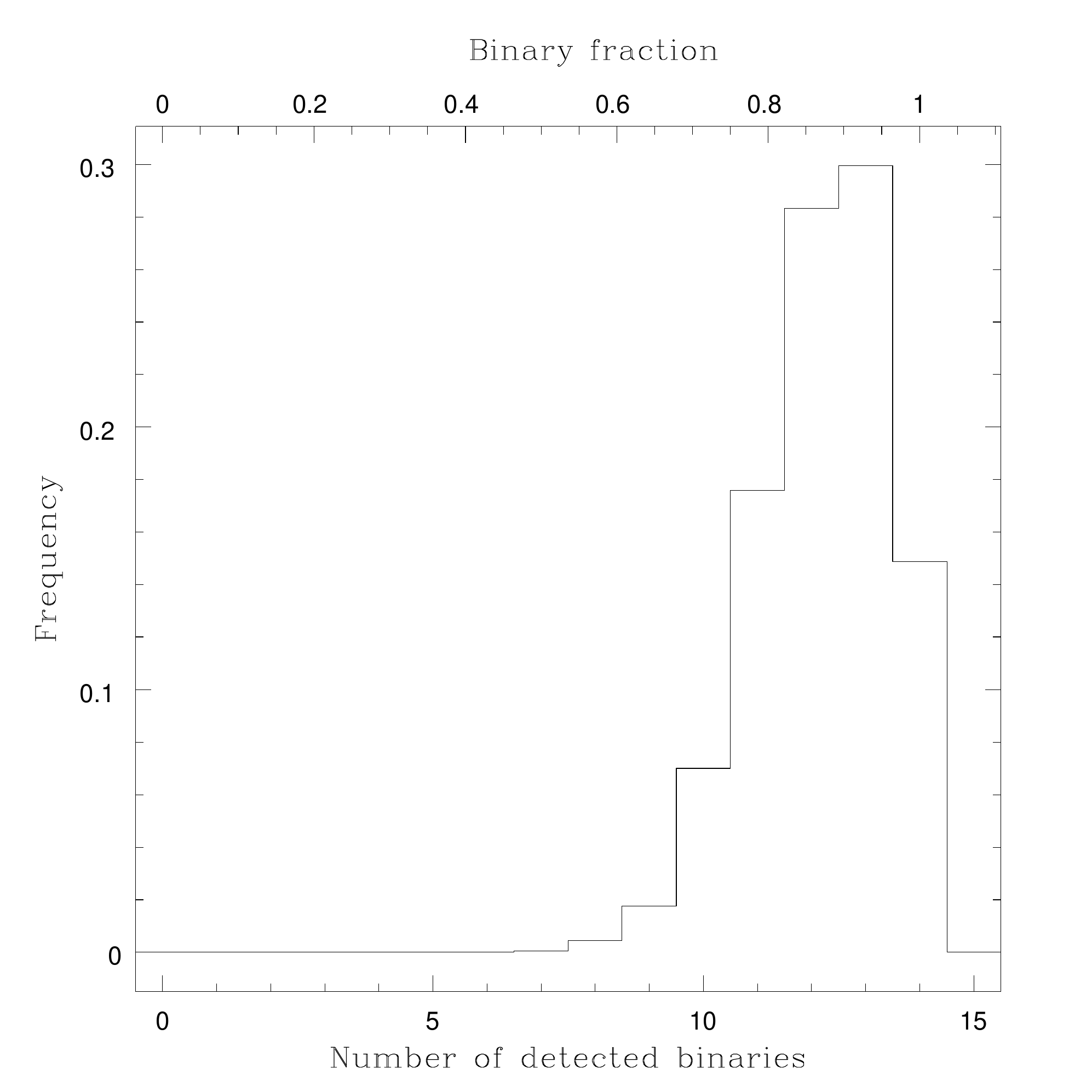}
\caption{Distribution of the number of detected SB systems as obtained from 10\,000 realizations of a cluster with 14 O-type binaries.}
\label{fig: bias}
\end{figure}
% ========================================================================

% Probability to detect 8 or less is 2.E-3, or 0.2%. ==> Rejection of the hypothesis of all being spectro. binaries with companion with q>0.2
% Binary fraction : 8/14: 57% (+/-13%)

\section{Observational biases} \label{sect: MC}

To estimate the impact of the observational biases on the detection probability of a binary with a given primary mass, we follow the approach described in \citetalias{SGE09}. Briefly, we first draw a series of $N$ SB2 systems adopting the following rules. The primary mass is fixed to the theoretical value of \citet{MSH05} according to the primary spectral type of the considered object. The probability density function (PDF) of $\log P$ (d) is taken to be uniform in the interval [0.3,1.0] and [1.0,3.5]. 50\%\ of the systems are spread in the first interval while the rest is distributed over the second interval. The eccentricity and mass-ratio PDFs are taken to be uniform in the range [0.0,0.9] and [0.1,1.0] respectively, with the additional constraint that the separation at periastron is larger than 20~\Rsun\ to avoid contact. Finally, the orbital configuration of each system is drawn assuming a random orientation in space and a random time of periastron passage. 

We then applied the time sampling corresponding to the observations of a given object to the RV curve of each of the $N$ SB2 binaries and we computed what would have been the largest RV difference ($\Delta RV$) recorded. The detection probability $P_\mathrm{detect}$ is then given by the ratio of the systems that display $\Delta RV$ larger than a given threshold $C$:
\begin{equation}
P_\mathrm{detect}=\frac{N(\Delta RV > C)}{N}.
\end{equation}
In the results presented in Table~\ref{tab: mc_indiv}, we adopted $C=20$~\kms\ and we computed $P_\mathrm{detect}$ for various ranges of the orbital periods. Our overall detection probability is better than 0.8 for all the objects, with the main uncertainties coming from the long period range. Based on those results one can exclude that the objects reported as single (Sect.~\ref{sect: sgl}) are short period binaries ($P<10$~d). Some of them could still be undetected long-period binaries. Yet, given that such systems are expected to be rare \citep{SaE10}, the chance of having at least one undetected system among the six presumably single objects is 0.56. 
% and remains thus not negligible. %but it is only 0.02 while the chance that all of them are undetected binaries is $3.5\times10^{-6}$.

Fig.~\ref{fig: bias} displays the probability density function of detecting a given number of binaries assuming all the objects in \ic\ are multiple systems.  It reveals that the probability to detect 8 binaries or less is 0.002, allowing thus to reject the null hypothesis of $f_\mathrm{true}=1.0$. This plot also indicates that the most likely binary detection rate of our campaign is close to 0.9.

 As discussed in \citetalias{SGE09}, our simulations neglect the effect of line blending, that decreases the detection probability of near equal flux systems and/or of systems with large projected rotational velocities. Among our sample, HD~308813 displays the broadest lines ($v \sin i \sim 200$\,\kms), yet RV variations with a peak-to-peak amplitude of $\sim40$\,\kms are clearly detected. All the presumably single stars discussed in Sect.~\ref{sect: sgl} have significantly narrower line profiles so that even small RV variations in an equal-flux system would induce significant changes in these line profiles. Finally, we note that the main limitations of our simulations lay in the adopted distributions of orbital parameters. While our assumptions are in line with earlier results from our campaign as well as with the preliminary results of \citet{SaE10}, it is unknown whether the multiplicity properties of O stars are homogeneous or show sizeable variations from one cluster to another.

%%%%%%%%%%%%%%%%%%%%%%%%%%%%%%%%%%%%%%%%%%%%%%%%%%%%%%%%%%%%%%%%%%%%%
%%%%%%%%%%%%%%%%%%%%%%%%%%%%%%%%%%%%%%%%%%%%%%%%%%%%%%%%%%%%%%%%%%%%%

% ========================================================================
\begin{table*}
 \centering
 \begin{minipage}{150mm}
  \caption{Final spectroscopic classification and multiplicity properties of the studied O stars in IC~2944. An 'O' appended to the SB1/2 flag means that an orbital solution is available; an 'E', that the system displays eclipses in addition to its SB signature.}
\label{tab: bin}
  \begin{tabular}{@{}llllllllll@{}}
  \hline
Object     & Mult.  & \multicolumn{2}{c}{Sp. Type}       & $M_2/M_1$    & $P$         & $e$          \\
           &        & Previous works    & This work      &                & (d)         &              \\
  \hline            
\multicolumn{7}{c}{SB2 systems} \\
  \hline                                          
HD~100099  & SB2O  & O9~III             & O9~III+O9.7~V   & 0.792          & 21.56       & 0.52         \\
HD~101131  & SB2O  & O6.5~V((f))+O8.5~V & \na            & 0.56           & 9.65        & 0.16         \\
HD~101190  & SB2O  & O5.5-O6~V((f))     & O4~V((f))+O7~V & 0.3-0.5        & 6.05        & $\approx0.3$ \\
HD~101205  & SB2OE & O7~IIIn((f))       & n/a            & undef.         & 2.1-2.8:    & 0.0:         \\
HD~101413  & SB2   & O8~V               & O8~V+B3:~V     & $\approx 0.2$  & $1-2\times10^2$ & undef.   \\ 
HD~101436  & SB2   & O6.5~V         & O6.5~V+O7~V     & 0.52-0.79$^a$      & 37.37        & 0.11    \\
  \hline            
\multicolumn{7}{c}{SB1 systems} \\
  \hline                               
HD~101191  & SB1   & O8~V((n))          & O8~V           & undef.         & $10^2-10^3$ & undef.       \\
HD~308804  & SB1   & O9.5~V             & B3~V           & undef.         & $>5$        & undef.       \\
HD~308813  & SB1   & O9.5~V-B0~V        & O9.5~V         & undef.         & undef.      & undef.       \\
  \hline            
\multicolumn{7}{c}{Single stars}\\
  \hline                               
HD~101223  & sgl.  & O8~V((f))          & O8~V           & \na & \na & \na  \\
HD~101298  & sgl.  & O6~III((f))        & O6~III((f))    & \na & \na & \na  \\
HD~101333  & sgl.  & O9.5-B0~III        & O9.5~V         & \na & \na & \na  \\
HD~101545A & sgl.  & O9.5-B0.5~Ib       & O9.5~III       & \na & \na & \na  \\
HD~101545B & sgl.  & O9.5-B0.5~Ib       & O9.7~III       & \na & \na & \na  \\
\cpd2198   & sgl.  & O9.5~Ib            & O9.7~III       & \na & \na & \na  \\
\hline
\end{tabular}\\
Notes: $a.$  For comparison purpose, we have inverted the primary and secondary identification so that the mass ratio is smaller than unity in the present table.
\end{minipage}
\end{table*}
% ========================================================================

% ========================================================================
\begin{figure}
\centering
\includegraphics[width=\columnwidth]{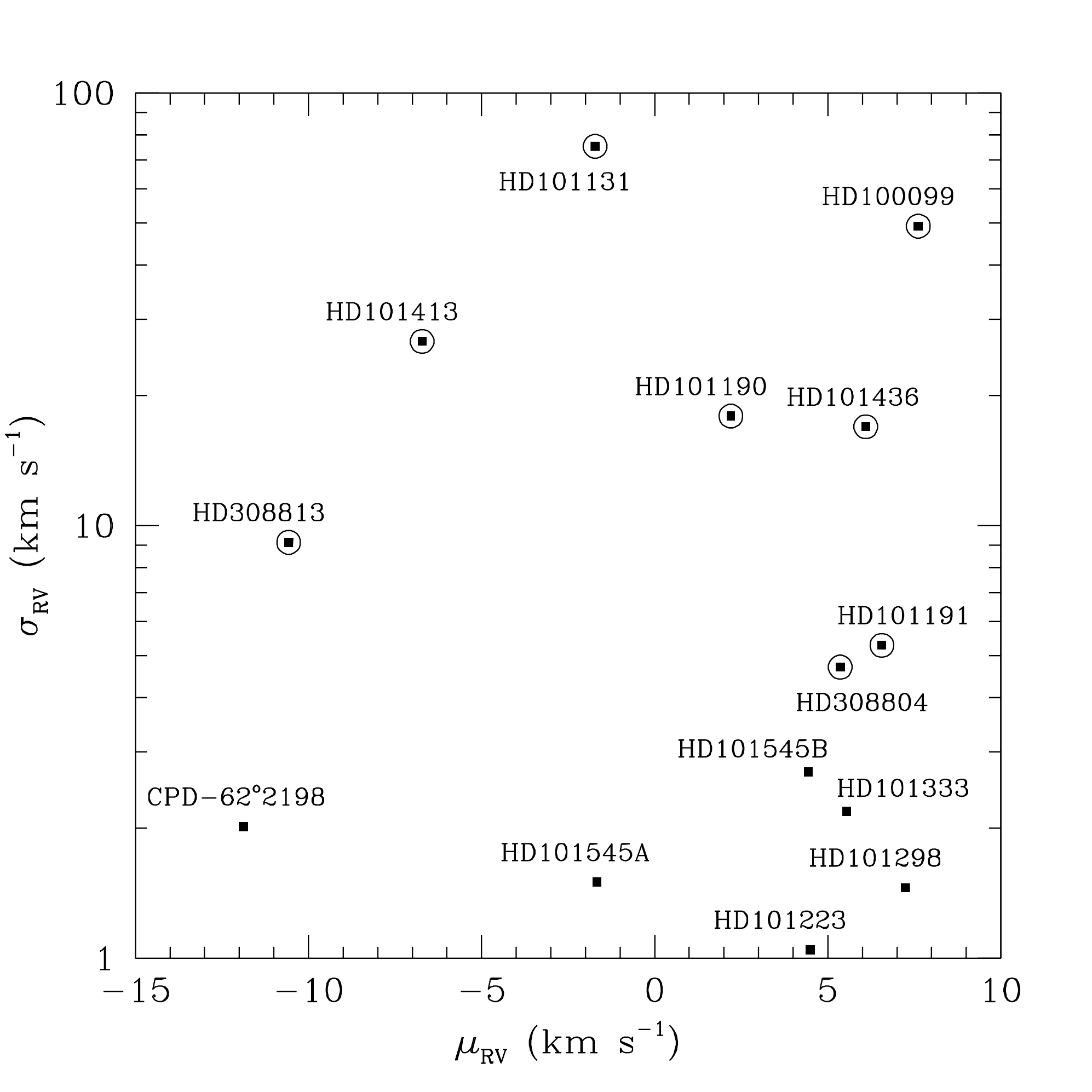}
\caption{Dispersion of the \hea\l5876 RV measurements versus their mean RV. Circles around the $\sigma-\mu$ points indicate the binaries.  The average systemic velocities have been used instead of the mean RV for the  binaries with orbital solutions : HD~100099, HD~101131,  HD~101190 and HD~101436.}
\label{fig: mu-sig}
\end{figure}
% ========================================================================

\section{Discussion} \label{sect: discuss}
% ========================================================================
\begin{table}
 \centering
 \begin{minipage}{80mm}
  \caption{Color excess and spectroscopic distance modulus (DM) computed from the $V$ and $K$ band magnitudes.}
\label{tab: dm}
  \begin{tabular}{@{}lccc@{}}
  \hline 
  Object & $E(\mathrm{B-V})$& $DM_\mathrm{V}$ & $DM_\mathrm{K}$  \\
         &                  &                & \\ 
  \hline   
  HD~100099   &    0.36  &  $12.57\pm 0.26$   &  $12.31 \pm 0.07$  \\  
  HD~101131   &    0.27  &  $11.68\pm 0.40$   &  $11.40 \pm 0.11$  \\  
  HD~101190   &    0.37  &  $12.08\pm 0.50$   &  $12.12 \pm 0.21$  \\  
  HD~101205   &    0.30  &  $11.43\pm 0.73$   &  $11.08 \pm 0.10$  \\  
  HD~101413   &    0.30  &  $11.86\pm 0.28$   &  $11.58 \pm 0.13$  \\  
  HD~101436   &    0.32  &  $12.14\pm 0.50$   &  $12.00 \pm 0.11$  \\  
  HD~101191   &    0.28  &  $12.03\pm 0.14$   &  $11.79 \pm 0.14$  \\  
  HD~308804   &    0.40  &  $11.73\pm 0.55$   &  $10.82 \pm 0.05$  \\  
  HD~308813   &    0.29  &  $12.28\pm 0.16$   &  $12.17 \pm 0.09$  \\  
  HD~101223   &    0.43  &  $11.77\pm 0.15$   &  $11.63 \pm 0.14$  \\  
  HD~101298   &    0.31  &  $12.86\pm 0.09$   &  $12.57 \pm 0.08$  \\  
  HD~101333   &    0.31  &  $12.02\pm 0.16$   &  $11.75 \pm 0.09$  \\  
  HD~101545AB &    0.23  &  $12.16\pm 0.32$   &  $11.40 \pm 0.05$  \\  
  \cpd2198   &    0.39  &  $14.46\pm 0.07$   &  $13.05 \pm 0.05$  \\  
  \hline 
\end{tabular}
\end{minipage}
\end{table}
% ========================================================================

\subsection{Distance and membership}\label{ssect: dist}

As already discussed in the Introduction, several authors questioned the single nature of the  \ic\ complex and preferred to rather consider several OB associations along the line of sight. Using our refined spectral classifications, we recompute the distance modulus (DM) of each target in our sample using both the $V$ and $K$ band magnitudes (Table~\ref{tab: dm}).   In this exercise, we adopted the theoretical absolute magnitudes, colors and bolometric corrections from \citet{MaP06} and the interstellar extinction law of \citet{Fit99}. 

The revision of the spectral classification and more particularly of the multiplicity status of the components leads to significant changes in the spectroscopic parallax of the objects. The most critical examples are probably the cases of HD~101191 and HD~101545 which are now located at a distance compatible with the other objects in IC~2944.   The largest deviations are shown by HD~308804 and \cpd2198 and these stars might respectively be foreground and background objects.  HD~101298 also seems slightly more distant. As noted earlier, adopting a main-sequence classification instead of a giant one would bring the star in  perfect agreement with the cluster mean distance. Ignoring these three stars as well as HD~101205 due to its uncertain number of components, 
 the estimated distance to IC~2944 is $2.3\pm0.3$~kpc, in perfect agreement with the earlier estimates of \citet{ThW65} and  \citet{TEH98},  but relatively larger than the 1.8~kpc estimate of \citet{KPR05}. More interestingly, we cannot identify significant differences between the various subgroups of stars in ~\ic, suggesting that most of the O stars belong to a single entity.

Finally, the dynamical properties of the objects compared to the average properties of the sample can also be used as a membership criterion. From Fig.~\ref{fig: mu-sig}, and beside the binaries, only CPD$-$62\degr2198 displays a significant RV difference compared to the systemic velocity of the cluster. Combined with the significantly different reddening and distance for the object (Table~\ref{tab: dm}), CPD$-$62\degr2198  potentially is a much more distant star seen along the line of sight to IC 2944.

\subsection{Binary fraction}

Eight objects out of 14 O-type stars show a clear SB signature, yielding a minimal binary fraction of $f_\mathrm{min}= 0.57$. Statistical uncertainties due to the limited size of our sample amounts to $\pm{0.13}$ (1\s\ error bar).  Using the results of Sect.~\ref{sect: MC}, the observed binary fraction likely corresponds to a true binary fraction $f_\mathrm{true}=0.65$. Proportionally, this makes this sample one of the richest binary nest after NGC~6231 \citep[][, Paper~I]{SGN08}. Yet, some objects might not belong to \ic\ itself. For example, HD~100099 and HD~101545AB are significantly offset from the cluster core and \cpd2198 seems much more distant than \ic. Rejecting these three objects, the SB fraction in \ic\ would  increase to 0.70.

Beside HD~101205, HD~101413 and HD~101545, \citet{MGH98} note single star results from speckle observations of six targets in Table~\ref{tab: bin}:  HD~100099, HD~101131, HD~101123, HD~101298, HD~101413, HD~101436.  Within a radius of 3\arcsec, the average number of companion per primary O star is thus of 0.84 and is again very similar to the value obtained in NGC~6231. As in the previous papers of this series, close to 70\%\ of the O star population is to be found in spectroscopic binary systems. 

\subsection{Orbital parameters}

Among the 8  spectroscopic binaries in our sample, three have periods up to 10 days, two others have periods of several tens of days, and the last three systems are longer period binaries with, most likely, an orbital time scale of the order of months to years. While the sample is likely too small to significantly constrain the overall period distribution, we note that there is no obvious contradiction with the period distribution derived for the binaries in NGC~6231 \citepalias{SGN08}, neither with the preliminary results of \citet{SaE10}. Similarly, the mass-ratio distribution is poorly sampled but would qualitatively follow the uniform distribution suggested by 
 \citet{SaE10}.

\section{Summary} \label{sect: ccl}

Using an extended set of high-resolution high SNR optical spectra, we revisited the status of 14 early-type objects in \ic\ and the Cen OB2 association. We presented new evidence of binarity for five objects and we confirmed the multiple nature of another two. We derived the first orbital solutions for HD~100099, HD~101436 and HD~101190 and provide additional support for a higher multiplicity in HD~101205. The minimal spectroscopic binary fraction in our sample is 0.57 but rises to 0.63 if one only considers the objects located in the inner 12\arcmin\ around HD~101205. Our sample is limited and does not provide strong  constraints on the distribution of the orbital parameters of the binary systems. Yet, no sizeable disagreement is observed with results from earlier papers in this series, nor with the preliminary results of \citet{SaE10} that combine data from several clusters. Using an approach similar to the one presented in \citetalias{SGE09}, we estimated the observational biases of our campaign and we showed that the binary detection rate is close to 90\%, leaving thus little room for undetected systems. Using newly derived spectroscopic parallaxes, we reassessed the distance to IC~2944. We confirm that, as far as the O stars are concerned, the cluster is most likely a single entity and that HD101545 and HD100099, offset by 50-60\arcmin\ from the cluster core, are located at a similar distance as IC~2944.

%Despite the fact that IC~2944 is the third richest concentration of O stars within 3~kpc (after the Carina region and NGC~6231), the investigation of the distribution of orbital parameters is still limited by small sample size effects. Yet, no strong discrepancy are found with results from previous papers in this series. 

%_____________ ACKNOWLEDGMENTS __________________________________________

\section*{Acknowledgments}
 This paper relies on data taken at the La Silla-Paranal Observatory under program IDs 073.D-0609(A), 075.D-0369(A), 077.D-0146(A) and 080.D-0855(A). The authors warmly thank the ESO staff for efficient support during both visitor and service mode runs and to the referee, Dr. D. Gies, for his constructive comments on the manuscript. EG is quite thankful to Jean-Philippe Beaulieu and to his friend for help and assistance
especially under the form of various tea infusions; he is also indebted to the ESO staff
and in particular to Monica Castillo for continuous and efficient support during a particularly harsh
run that ended at Clinica del Elqui.
This work made use of the SIMBAD and WEBDA databases and of the Vizier catalogue access tool (CDS, Strasbourg, France).

%_____________ BIBLIOGRAPHY _____________________________________________

\bibliographystyle{mn2e}
\bibliography{/home/hsana/Desktop/Dropbox/literature}

\begin{thebibliography}{}

\bibitem[\protect\citeauthoryear{{Alter}, {Balazs}, {Ruprecht} \&
  {Vanysek}}{{Alter} et~al.}{1970}]{ABR70}
{Alter} G.,  {Balazs} B.,  {Ruprecht} J.,    {Vanysek} J.,  1970, Catalogue of
  star clusters and associations.
Akademiai Kiado, Budapest

\bibitem[\protect\citeauthoryear{{Ardeberg} \& {Maurice}}{{Ardeberg} \&
  {Maurice}}{1977}]{ArM77}
{Ardeberg} A.,  {Maurice} E.,  1977, \aaps, 28, 153

\bibitem[\protect\citeauthoryear{{Ardeberg} \& {Maurice}}{{Ardeberg} \&
  {Maurice}}{1980}]{ArM80}
{Ardeberg} A.,  {Maurice} E.,  1980, \aaps, 39, 325

\bibitem[\protect\citeauthoryear{{Balona}}{{Balona}}{1992}]{Bal92}
{Balona} L.~A.,  1992, \mnras, 254, 404

\bibitem[\protect\citeauthoryear{{Baumgardt}, {Dettbarn} \&
  {Wielen}}{{Baumgardt} et~al.}{2000}]{BDW00}
{Baumgardt} H.,  {Dettbarn} C.,    {Wielen} R.,  2000, \aaps, 146, 251

\bibitem[\protect\citeauthoryear{{Charbonneau}}{{Charbonneau}}{1995}]{Cha95}
{Charbonneau} P.,  1995, \apjs, 101, 309

\bibitem[\protect\citeauthoryear{{Collinder}}{{Collinder}}{1931}]{Col31}
{Collinder} P.,  1931, Annals of the Observatory of Lund, 2, 1

\bibitem[\protect\citeauthoryear{{Conti}}{{Conti}}{1973}]{Con73_teff}
{Conti} P.~S.,  1973, \apj, 179, 181

\bibitem[\protect\citeauthoryear{{Conti} \& {Alschuler}}{{Conti} \&
  {Alschuler}}{1971}]{Ca71}
{Conti} P.~S.,  {Alschuler} W.~R.,  1971, \apj, 170, 325

\bibitem[\protect\citeauthoryear{{Conti}, {Leep} \& {Lorre}}{{Conti}
  et~al.}{1977}]{CLL77}
{Conti} P.~S.,  {Leep} E.~M.,    {Lorre} J.~J.,  1977, \apj, 214, 759

\bibitem[\protect\citeauthoryear{{Fitzpatrick}}{{Fitzpatrick}}{1999}]{Fit99}
{Fitzpatrick} E.~L.,  1999, \pasp, 111, 63

\bibitem[\protect\citeauthoryear{{Fran{\c c}ois}, {Depagne}, {Hill}, {Spite},
  {Spite}, {Plez}, {Beers}, {Andersen}, {James}, {Barbuy}, {Cayrel},
  {Bonifacio}, {Molaro}, {Nordstr{\"o}m} \& {Primas}}{{Fran{\c c}ois}
  et~al.}{2007}]{FDH07}
{Fran{\c c}ois} P.,  {Depagne} E.,  {Hill} V.,  {Spite} M.,  {Spite} F.,
  {Plez} B.,  {Beers} T.~C.,  {Andersen} J.,  {James} G.,  {Barbuy} B.,
  {Cayrel} R.,  {Bonifacio} P.,  {Molaro} P.,  {Nordstr{\"o}m} B.,    {Primas}
  F.,  2007, \aap, 476, 935

\bibitem[\protect\citeauthoryear{{Garrison}, {Hiltner} \& {Schild}}{{Garrison}
  et~al.}{1977}]{GHS77}
{Garrison} R.~F.,  {Hiltner} W.~A.,    {Schild} R.~E.,  1977, \apjs, 35, 111

\bibitem[\protect\citeauthoryear{{Gies}, {Penny}, {Mayer}, {Drechsel} \&
  {Lorenz}}{{Gies} et~al.}{2002}]{GPM02}
{Gies} D.~R.,  {Penny} L.~R.,  {Mayer} P.,  {Drechsel} H.,    {Lorenz} R.,
  2002, \apj, 574, 957

\bibitem[\protect\citeauthoryear{{Gosset}, {Royer}, {Rauw}, {Manfroid} \&
  {Vreux}}{{Gosset} et~al.}{2001}]{GRR01}
{Gosset} E.,  {Royer} P.,  {Rauw} G.,  {Manfroid} J.,    {Vreux} J.-M.,  2001,
  \mnras, 327, 435

\bibitem[\protect\citeauthoryear{{Heck}, {Manfroid} \& {Mersch}}{{Heck}
  et~al.}{1985}]{HMM85}
{Heck} A.,  {Manfroid} J.,    {Mersch} G.,  1985, \aaps, 59, 63

\bibitem[\protect\citeauthoryear{{Herbst}}{{Herbst}}{1975}]{Her75}
{Herbst} W.,  1975, \aj, 80, 212

\bibitem[\protect\citeauthoryear{{Howarth}, {Siebert}, {Hussain} \&
  {Prinja}}{{Howarth} et~al.}{1997}]{HSHP97}
{Howarth} I.~D.,  {Siebert} K.~W.,  {Hussain} G.~A.~J.,    {Prinja} R.~K.,
  1997, \mnras, 284, 265

\bibitem[\protect\citeauthoryear{{Huang} \& {Gies}}{{Huang} \&
  {Gies}}{2006a}]{HuG06a}
{Huang} W.,  {Gies} D.~R.,  2006a, \apj, 648, 580

\bibitem[\protect\citeauthoryear{{Huang} \& {Gies}}{{Huang} \&
  {Gies}}{2006b}]{HuG06b}
{Huang} W.,  {Gies} D.~R.,  2006b, \apj, 648, 591

\bibitem[\protect\citeauthoryear{{Humphreys}}{{Humphreys}}{1973}]{Hum73}
{Humphreys} R.~M.,  1973, \aaps, 9, 85

\bibitem[\protect\citeauthoryear{{James}, {Fran{\c c}ois}, {Bonifacio},
  {Carretta}, {Gratton} \& {Spite}}{{James} et~al.}{2004}]{JFB04}
{James} G.,  {Fran{\c c}ois} P.,  {Bonifacio} P.,  {Carretta} E.,  {Gratton}
  R.~G.,    {Spite} F.,  2004, \aap, 427, 825

\bibitem[\protect\citeauthoryear{{Kharchenko}, {Piskunov}, {R{\"o}ser},
  {Schilbach} \& {Scholz}}{{Kharchenko} et~al.}{2005}]{KPR05}
{Kharchenko} N.~V.,  {Piskunov} A.~E.,  {R{\"o}ser} S.,  {Schilbach} E.,
  {Scholz} R.-D.,  2005, \aap, 438, 1163

\bibitem[\protect\citeauthoryear{{Martins} \& {Plez}}{{Martins} \&
  {Plez}}{2006}]{MaP06}
{Martins} F.,  {Plez} B.,  2006, \aap, 457, 637

\bibitem[\protect\citeauthoryear{{Martins}, {Schaerer} \& {Hillier}}{{Martins}
  et~al.}{2005}]{MSH05}
{Martins} F.,  {Schaerer} D.,    {Hillier} D.~J.,  2005, \aap, 436, 1049

\bibitem[\protect\citeauthoryear{{Mason}, {Gies}, {Hartkopf}, {Bagnuolo}, {ten
  Brummelaar} \& {McAlister}}{{Mason} et~al.}{1998}]{MGH98}
{Mason} B.~D.,  {Gies} D.~R.,  {Hartkopf} W.~I.,  {Bagnuolo} W.~G.,  {ten
  Brummelaar} T.,    {McAlister} H.~A.,  1998, \aj, 115, 821

\bibitem[\protect\citeauthoryear{{Mathys}}{{Mathys}}{1988}]{Mat88}
{Mathys} G.,  1988, \aaps, 76, 427

\bibitem[\protect\citeauthoryear{{Mathys}}{{Mathys}}{1989}]{Mat89}
{Mathys} G.,  1989, \aaps, 81, 237

\bibitem[\protect\citeauthoryear{{Mayer}, {Bo{\v z}i{\'c}}, {Lorenz} \&
  {Drechsel}}{{Mayer} et~al.}{2010}]{MBL10}
{Mayer} P.,  {Bo{\v z}i{\'c}} H.,  {Lorenz} R.,    {Drechsel} H.,  2010,
  Astronomische Nachrichten, 331, 274

\bibitem[\protect\citeauthoryear{{Mayer}, {Lorenz} \& {Drechsel}}{{Mayer}
  et~al.}{1992}]{MLD92}
{Mayer} P.,  {Lorenz} R.,    {Drechsel} H.,  1992, Information Bulletin on
  Variable Stars, 3765, 1

\bibitem[\protect\citeauthoryear{{Mermilliod}}{{Mermilliod}}{1992}]{Me92}
{Mermilliod} J.-C.,  1992, Bulletin d'Information du Centre de Donn\'ees
  Stellaires (CDS), 40, 115

\bibitem[\protect\citeauthoryear{{Otero}}{{Otero}}{2007}]{Ote07}
{Otero} S.~A.,  2007, Open European Journal on Variable Stars, 72, 1

\bibitem[\protect\citeauthoryear{{Penny}}{{Penny}}{1996}]{Pen96}
{Penny} L.~R.,  1996, \apj, 463, 737

\bibitem[\protect\citeauthoryear{{Penny} \& {Gies}}{{Penny} \&
  {Gies}}{2009}]{PeG09}
{Penny} L.~R.,  {Gies} D.~R.,  2009, \apj, 700, 844

\bibitem[\protect\citeauthoryear{{Perry} \& {Landolt}}{{Perry} \&
  {Landolt}}{1986}]{PeL86}
{Perry} C.~L.,  {Landolt} A.~U.,  1986, \aj, 92, 844

\bibitem[\protect\citeauthoryear{{Puls}, {Urbaneja}, {Venero}, {Repolust},
  {Springmann}, {Jokuthy} \& {Mokiem}}{{Puls} et~al.}{2005}]{PUV05}
{Puls} J.,  {Urbaneja} M.~A.,  {Venero} R.,  {Repolust} T.,  {Springmann} U.,
  {Jokuthy} A.,    {Mokiem} M.~R.,  2005, \aap, 435, 669

\bibitem[\protect\citeauthoryear{{Reed} \& {Kuhna}}{{Reed} \&
  {Kuhna}}{1997}]{ReK97}
{Reed} B.~C.,  {Kuhna} K.~M.,  1997, \aj, 113, 823

\bibitem[\protect\citeauthoryear{{Reipurth}, {Corporon}, {Olberg} \&
  {Tenorio-Tagle}}{{Reipurth} et~al.}{1997}]{RCO97}
{Reipurth} B.,  {Corporon} P.,  {Olberg} M.,    {Tenorio-Tagle} G.,  1997,
  \aap, 327, 1185

\bibitem[\protect\citeauthoryear{{Reipurth}, {Raga} \& {Heathcote}}{{Reipurth}
  et~al.}{2003}]{RRH03}
{Reipurth} B.,  {Raga} A.,    {Heathcote} S.,  2003, \aj, 126, 1925

\bibitem[\protect\citeauthoryear{{Sana} \& {Evans}}{{Sana} \&
  {Evans}}{2010}]{SaE10}
{Sana} H.,  {Evans} C.,  2010, in {C.~Neiner, G.~Wade, G.~Meynet \& G. ~Peters}
  ed., {Active OB stars} Vol.~272 of IAUS, {The multiplicity of massive stars}.
Cambridge University Press, in press (arXiv: 1009.4197)

\bibitem[\protect\citeauthoryear{{Sana}, {Gosset} \& {Evans}}{{Sana}
  et~al.}{2009}]{SGE09}
{Sana} H.,  {Gosset} E.,    {Evans} C.~J.,  2009, \mnras, 400, 1479

\bibitem[\protect\citeauthoryear{{Sana}, {Gosset}, {Naz{\'e}}, {Rauw} \&
  {Linder}}{{Sana} et~al.}{2008}]{SGN08}
{Sana} H.,  {Gosset} E.,  {Naz{\'e}} Y.,  {Rauw} G.,    {Linder} N.,  2008,
  \mnras, 386, 447

\bibitem[\protect\citeauthoryear{{Sana}, {Gosset} \& {Rauw}}{{Sana}
  et~al.}{2006}]{SGR06_219}
{Sana} H.,  {Gosset} E.,    {Rauw} G.,  2006, \mnras, 371, 67

\bibitem[\protect\citeauthoryear{{Sana}, {Rauw} \& {Gosset}}{{Sana}
  et~al.}{2001}]{SRG01}
{Sana} H.,  {Rauw} G.,    {Gosset} E.,  2001, \aap, 370, 121

\bibitem[\protect\citeauthoryear{{Schild}}{{Schild}}{1970}]{Sch70}
{Schild} R.~E.,  1970, \apj, 161, 855

\bibitem[\protect\citeauthoryear{{Thackeray}}{{Thackeray}}{1950}]{Tha50}
{Thackeray} A.~D.,  1950, \mnras, 110, 524

\bibitem[\protect\citeauthoryear{{Thackeray} \& {Wesselink}}{{Thackeray} \&
  {Wesselink}}{1965}]{ThW65}
{Thackeray} A.~D.,  {Wesselink} A.~J.,  1965, \mnras, 131, 121

\bibitem[\protect\citeauthoryear{{Tovmassian}, {Epremian}, {Hovhannessian},
  {Cruz-Gonzalez}, {Navarro} \& {Karapetian}}{{Tovmassian}
  et~al.}{1998}]{TEH98}
{Tovmassian} H.~M.,  {Epremian} R.~A.,  {Hovhannessian} K.,  {Cruz-Gonzalez}
  G.,  {Navarro} S.~G.,    {Karapetian} A.~A.,  1998, \aj, 115, 1083

\bibitem[\protect\citeauthoryear{{Tovmassian}, {Khodzhayants}, {Krmoyan},
  {Kashin}, {Zakharian}, {Oganessian}, {Mkrtchian}, {Tovmassian}, {Huguenin},
  {Butov}, {Romanenko}, {Laveykin} \& {Aleksandrov}}{{Tovmassian}
  et~al.}{1988}]{TKK88}
{Tovmassian} H.~M.,  {Khodzhayants} Y.~M.,  {Krmoyan} M.~N.,  {Kashin} A.~L.,
  {Zakharian} A.~Z.,  {Oganessian} R.~K.,  {Mkrtchian} M.~A.,  {Tovmassian}
  G.~H.,  {Huguenin} D.,  {Butov} V.~V.,  {Romanenko} Y.~B.,  {Laveykin} A.~I.,
     {Aleksandrov} A.~P.,  1988, Pis'ma v Astronomicheskii Zhurnal, 14, 291

\bibitem[\protect\citeauthoryear{{Vega}, {Orsatti} \& {Marraco}}{{Vega}
  et~al.}{1994}]{VOM94}
{Vega} E.~I.,  {Orsatti} A.~M.,    {Marraco} H.~G.,  1994, \aj, 108, 1834

\bibitem[\protect\citeauthoryear{{Walborn}}{{Walborn}}{1973}]{Wal73}
{Walborn} N.~R.,  1973, \aj, 78, 1067

\bibitem[\protect\citeauthoryear{{Walborn}}{{Walborn}}{1987}]{Wal87}
{Walborn} N.~R.,  1987, \aj, 93, 868

\bibitem[\protect\citeauthoryear{{Walborn} \& {Fitzpatrick}}{{Walborn} \&
  {Fitzpatrick}}{1990}]{WF90}
{Walborn} N.~R.,  {Fitzpatrick} E.~L.,  1990, \pasp, 102, 379

\end{thebibliography}

\bsp

\label{lastpage}

\end{document}